\documentclass[twocolumn]{aastex631}
\usepackage{amsmath}
\usepackage{enumitem}

\begin{document}

\title{The companion mass distribution of post common envelope hot subdwarf binaries: evidence for boosted and disrupted magnetic braking?}

\author{Lisa Blomberg}
\affiliation{Department of Astronomy, California Institute of Technology, 1200 East California Boulevard, Pasadena, CA 91125, USA}
\author{Kareem El-Badry}
\affiliation{Department of Astronomy, California Institute of Technology, 1200 East California Boulevard, Pasadena, CA 91125, USA}
\author{Katelyn Breivik}
\affiliation{McWilliams Center for Cosmology, Department of Physics, Carnegie Mellon University, Pittsburgh, PA 15213, USA}
\author{Ilaria Caiazzo}
\affiliation{Institute of Science and Technology Austria, Am Campus 1, 3400 Klosterneuburg, Austria}
\author{Pranav Nagarajan}
\affiliation{Department of Astronomy, California Institute of Technology, 1200 East California Boulevard, Pasadena, CA 91125, USA}
\author{Antonio Rodriguez}
\affiliation{Department of Astronomy, California Institute of Technology, 1200 East California Boulevard, Pasadena, CA 91125, USA}
\author{Jan van Roestel}
\affiliation{Anton Pannekoek Institute for Astronomy, University of Amsterdam, 1090 GE Amsterdam, The Netherlands
3}
\author{Zachary P. Vanderbosch}
\affiliation{Department of Astronomy, California Institute of Technology, 1200 East California Boulevard, Pasadena, CA 91125, USA}
\author{Natsuko Yamaguchi}
\affiliation{Department of Astronomy, California Institute of Technology, 1200 East California Boulevard, Pasadena, CA 91125, USA}





\begin{abstract}

We measure the mass distribution of main-sequence (MS) companions to hot subdwarf B stars (sdBs) in post-common envelope binaries (PCEBs). We carried out a spectroscopic survey of 14 eclipsing systems (``HW Vir binaries'') with orbital periods of $3.8 < P_{\rm orb} < 12$ hours, resulting in a well-understood selection function and a near-complete sample of HW Vir binaries with $G < 16$. We constrain companion masses from the radial velocity curves of the sdB stars. The companion mass distribution peaks at $M_{\rm MS}\approx 0.15 M_{\odot}$ and drops off at $M_{\rm MS} > 0.2\,M_{\odot}$, with only two systems hosting companions above the fully-convective limit. There is no correlation between $P_{\rm orb}$ and $M_{\rm MS}$ within the sample. A similar drop-off in the companion mass distribution of white dwarf (WD) + MS PCEBs has been attributed to disrupted magnetic braking (MB) below the fully-convective limit. We compare the sdB companion mass distribution to predictions of binary evolution simulations with a range of MB laws. Because sdBs have short lifetimes compared to WDs, explaining the lack of higher-mass MS companions to sdBs with disrupted MB requires MB to be boosted by a factor of 20-100 relative to MB laws inferred from the rotation evolution of single stars. We speculate that such boosting may be a result of irradiation-driven enhancement of the MS stars' winds. An alternative possibility is that common envelope evolution favors low-mass companions in short-period orbits, but the existence of massive WD companions to sdBs with similar periods disfavors this scenario.


\end{abstract}



\section{Introduction} \label{sec:intro}
Magnetic braking (MB) is one of the primary drivers of angular momentum loss in binaries. Magnetic fields cause mass lost in a stellar wind to co-rotate with the star, such that the angular momentum of the wind is greater than that of the stellar surface. Over time, this removes angular momentum and causes the star to spin down \citep[e.g.][]{Schatzman_1962, Weber_1967, Mestal_1968, Mestel_1987}.

However, in close binaries, tides synchronize the component stars' rotation with the orbital period \citep[e.g.][]{Zahn_1977}. As a result, MB is not able to spin down either of the components. Instead, the wind removes angular momentum from the orbit, causing it to shrink and the components to spin up. This process plays a critical role in the evolution of close binaries: it brings detached binaries into contact and sets the period evolution and mass transfer rates of mass-transfer binaries, such as cataclysmic variables \citep[e.g.][]{Knigge_2011}. However, MB is imperfectly understood theoretically. Most MB models used in binary evolution calculations are empirically calibrated from observations of the spin evolution of single stars \citep[e.g.][]{Skumanich_1972, Matt_2015}, most of which rotate slower than the stars in close, tidally-locked binaries.

The MB prescription most widely used in the binary evolution literature is the model proposed by \cite{Verbunt_Zwaan_1981} and refined by \citet[][hereafter RVJ]{RVJ_1983}. This model predicts a MB torque that scales as ${P_{\rm orb}}^{-3}$, leading to accelerated inspiral at short periods. A key feature of this model is the prediction that MB is ``disrupted'' when a star becomes fully convective. This disruption was proposed to explain the ``period gap'' in the orbital period distribution of cataclysmic variables (CVs) at $2-3$ hours, corresponding to donor masses near the fully-convective boundary. In the disrupted magnetic braking (DMB) paradigm, this gap occurs because MB abruptly weakens when CV donors become fully convective, causing the donors to return to thermal equilibrium and temporarily halting mass transfer. 

In support of the DMB model, it has been observed that the mass distribution of the main sequence (MS) stars in close detached white dwarf (WD) + MS binaries peaks near $\sim 0.3 M_\odot$ and falls off steeply at higher masses. A seminal study by \cite{Schreiber_2010} carried out multi-epoch radial velocity (RV) follow-up on the published sample of WD + MS binaries from the Sloan Digital Sky Survey 
\citep[SDSS;][]{Rebassa-Mansergas_2010}. \cite{Politano_2006} had previously proposed a test of the DMB paradigm based on the post common envelope binary (PCEB) companion mass distribution, predicting that the fraction of PCEBs relative to wide WD + MS binaries should decrease by 37-73\% above $M_{\rm MS}\sim0.37 M_\odot$, which was their estimate of the fully convective limit. This drop is predicted because in the DMB paradigm, PCEBs with MS stars above the fully convective limit undergo MB-driven orbital inspiral and come into contact faster than PCEBs with fully-convective MS stars. \cite{Schreiber_2010} measured an 80\% drop in the fraction of PCEBs above $M_{\rm MS} \sim 0.3 M_\odot$, interpreting this as evidence of DMB. As we discuss below, significant uncertainties in the MB law and its role in driving orbital evolution of PCEBs remain, motivating us to revisit the PCEB test of the DMB model.

Several results suggest that MB in low-mass stars may be more complex than suggested by the RVJ model. First, studies of the rotational evolution of single low-mass stars in clusters have found spin rates to vary with age in a manner that cannot be explained with simple prescriptions depending only on the structural parameters of the stars \citep[e.g.][]{Brown_2014, Newton_2016, Bouma_2023}. In fact, some studies have even reported evidence for an {\it increase} in MB below the fully-convective limit \citep{Lu_2024}. Second, there is now a large body of evidence that MB torques do not scale as ${P_{\rm orb}}^{-3}$ at arbitrarily fast rotation rates, but instead saturate above a critical rotation rate corresponding to periods of order 10 days and thereafter have a shallower scaling with $P_{\rm orb}$ \citep[e.g.][]{Reiners_2009, Matt_2015, el-badry_2022}. Third, several studies have proposed that CVs experience additional angular momentum losses as a consequence of the mass transfer process, in addition to those associated with MB \citep[e.g.][]{Schreiber_2010}.

Recently, \cite{belloni_2023} investigated whether a saturated MB model -- which predicts that MB torques saturated above a critical rotation rate and is more consistent with observations of the rotation evolution of single stars and with the period distribution of detached MS + MS binaries -- could also explain the companion mass distribution in the SDSS WD + MS PCEB sample when combined with disruption of MB below the fully convective limit. They found that it could, but only if the MB torque is at least $\sim50$ times stronger than predicted by the models calibrated to single-star rotation rates. 

While \cite{Schreiber_2010} and \cite{belloni_2023} showed that the mass distribution of MS stars in WD + MS PCEBs could be explained as a result of MB, this may not be the only possible explanation, since the initial mass distribution of PCEBs directly after common envelope is uncertain.

In this paper, we investigate the companion mass distribution of another population of PCEBs: those containing a stripped core helium burning star \citep[an ``sdB'', e.g.][]{Heber_2016} and a MS star. We focus on eclipsing systems, also known as HW Vir binaries. Close sdB + MS binaries are formed when a binary containing a red giant and a MS star undergoes a common envelope event during or shortly before the giant's core helium flash. If the MS star's orbital inspiral releases enough energy to eject the envelope, we are left with a detached binary comprised of a naked helium-burning star (i.e., the sdB) and a MS star. Close sdB + MS binaries thus form through a very similar process to WD + MS PCEBs. The key difference are that (a) the common envelope interaction in sdB + MS binaries must have occurred near the tip of the giant branch \citep[e.g.][]{Han_2002} and (b) it must have occurred relatively recently, since sdB stars only live for about 100 Myr \citep[][]{Dorman_1993, Schindler_2015}. Thus, there is less time for MB to shrink the orbit during the lifetime of the sdB, and we expect the period and mass distribution of sdB + MS binaries to more closely reflect the distribution immediately after common envelope. 

If the observed lack of MS stars above the fully convective limit in WD + MS binaries is a result of DMB, we should expect a weaker drop-off toward higher masses in the mass distribution of sdB companions. Our basic approach is to measure the companion mass distribution of an observed sample of sdB + MS binaries with a well-understood selection function and then compare to predictions of population synthesis calculations carried out with a variety of MB prescriptions. Our analysis is carried out on a sample selected and spectroscopically followed-up specifically with the goal of constraining the companion mass distribution, allowing us to account and correct for selection biases when comparing models to data.

The rest of the paper is organized as follows. In Section \ref{sec: target selection}, we discuss our target selection and basic properties of the sample. In Section \ref{sec: follow-up observations}, we describe our spectroscopic observations. Then in Section \ref{sec: analysis}, we describe measurement of radial velocities, orbit fits, and companion mass measurements. In Section \ref{sec: comparing observations to simulated populations}, we  compare our observational results to the predictions of binary population synthesis models with a range of MB laws. We also discuss our results in the context of previous literature on WD + MS binaries. Finally, we conclude with a discussion of our findings and their implications in Section \ref{sec: conclusion}. 

\section{Observed sample}\label{sec: target selection}
\subsection{Sample selection}
Although there have been several previous population studies of sdB + MS systems \citep[e.g.][]{Kupfer_2015, Lei_2020, Kruckow_2021, Schaffenroth_2022}, these all consisted of heterogeneous population samples and were dominated by short-period binaries. Here, we construct a carefully selected sample with a well-understood selection function. We selected our targets from the list of sdB + MS binaries published by \citet{Schaffenroth_2022}, who conducted an all-sky search for sdBs in close binaries using light curves from the Transiting Exoplanet Survey Satellite \citep[TESS;][]{TESS}. We focus on eclipsing systems because they are close to edge-on, meaning there is little inclination uncertainty, and we can measure the companion mass from radial velocities (RVs). Additionally, the detection probability of these eclipsing systems 
depends primarily on geometry, making it easier to model the selection function. 

\cite{Schaffenroth_2022} built a sample of 52 HW Vir systems by drawing from the \cite{Geier2020} catalog of sdBs. Of these, we selected 37 systems that are brighter than $G = 16$ mag. Given the large amplitude of the reflection effect and eclipses in close sdB + MS binaries, the \cite{Schaffenroth_2022} sample is essentially complete to sdB + MS binaries with $G < 16$ mag and $P_{\rm orb}<1$ day, if they are in the \citet{Geier2020} catalog. Of these, we selected the 16 systems with periods between $3.8 - 12$ hours. We do not include shorter period systems because those systems can only host low-mass companions ($M_{\rm MS} \lesssim 0.3 M_\odot$). Of these 16, one system (J0531-6953) has previously been studied in detail by \cite{Kupfer_2015}. Out of the remaining 15 systems, we obtained multi-epoch spectroscopic follow-up of 13 (see Table \ref{table: follow-up}). Our final sample of 14 binaries is 87.5 $\%$ complete with respect to the \cite{Geier2020} sample of HW Vir systems brighter than $G = 16$ mag with periods of $3.8 - 12$ hours. The \citet{Geier2020} sample may not be complete in this magnitude range -- it is likely to be missing some sdBs in regions of high extinction -- but we show in Section~\ref{sec:observables} that the catalog's completeness is expected to be independent of companion mass over the mass range relevant to our study.

\subsection{Basic properties of the sample}
\subsubsection{Orbital ephemerides} \label{sec: period and eclipse time}
We now investigate the 14 systems in our sample in detail. We begin by measuring orbital periods and ephemerides from their light curves. 
 
We obtained light curves for each target from either the Zwicky Transient Facility \citep[ZTF;][]{ZTF} or the All-Sky Automated Survey for Supernovae \citep[ASASSN;][]{ASSASN_lightcurve_1}. We used light curves from ZTF and ASASSN rather than TESS because they have longer time baselines ($5-10$ years, compared to $\sim27$ days per TESS sector), allowing for more precise period measurements. When possible, we used the ZTF $r$-band light curve. For targets with poor ZTF phase coverage, we used the ASASSN $g$-band light curves.

\begin{figure*}[ht!]
\includegraphics[width = 0.94\textwidth]{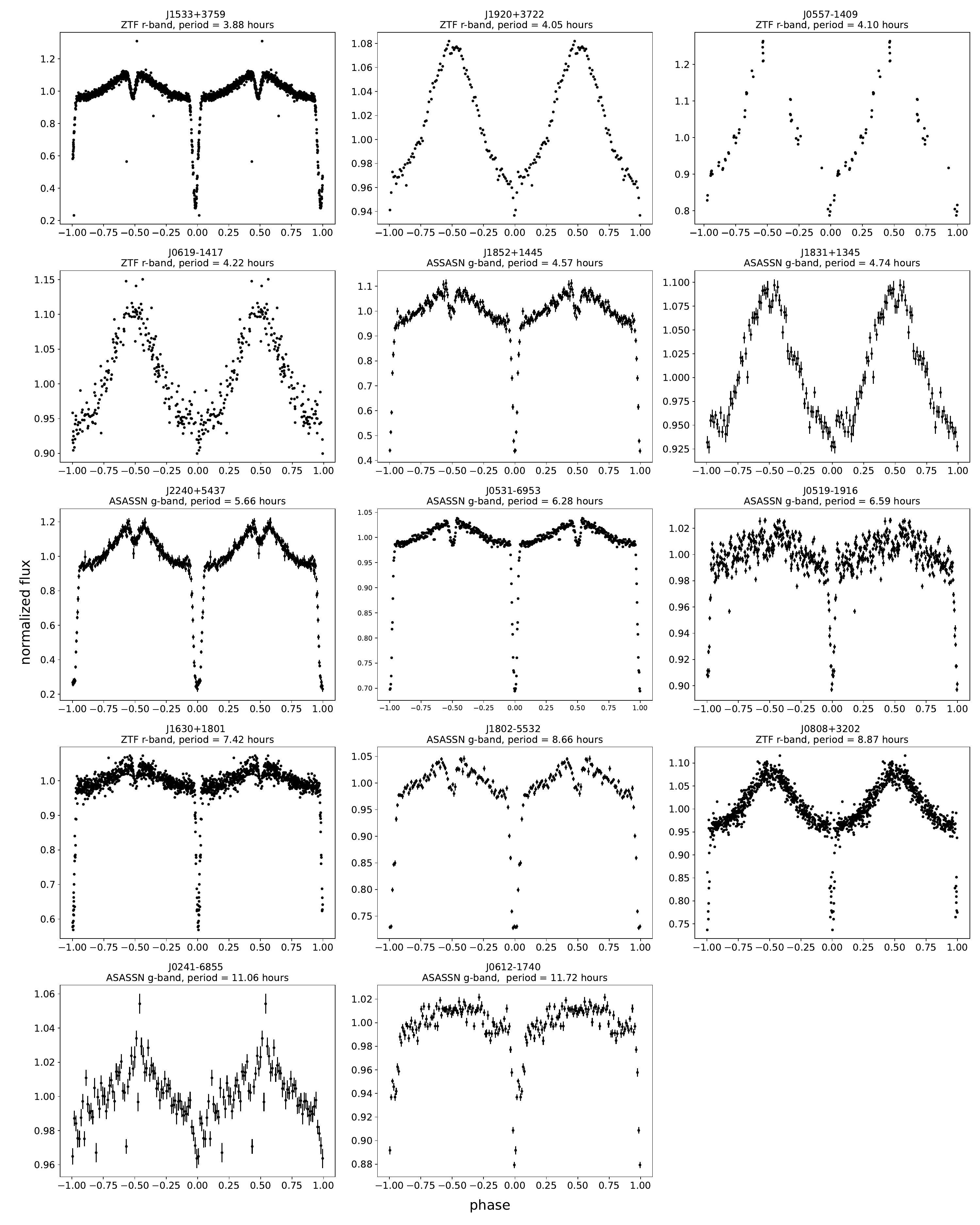}
\caption{Phased ZTF and ASASSN light curves for all targets in our sample, organized from shortest to longest period. We define phase 0 as the time of the primary eclipse (Table \ref{table: period and t_0}). A secondary eclipse is also visible in most systems. Data are from ZTF and ASASSN, and the light curves are phases-folded to the periods we determined (see Table \ref{table:basic_properties}).}
\label{fig: lightcurves}
\end{figure*}

We calculated a Lomb-Scargle periodogram \citep{Lomb_1976,Scargle_1982} using the Python Astropy package \citep{astropy:2013, astropy:2018, astropy:2022} to determine orbital periods. Phased and normalized light curves for each source are shown in Figure \ref{fig: lightcurves}, where we also indicate which survey and bandpass was used for each system. For several systems which exhibited noisy light curves after phase folding, we binned the photometric data. We binned our data into $N=75, 100, 150$ or $300$ bins, where we averaged over the fluxes within each bins. The photometric variability is dominated by quasi-sinusoidal variability on the orbital period, which is a result of irradiation of the ``day'' side of the MS star by the sdB. Primary and secondary eclipses are also evident for all sources, confirming their nature as HW Vir binaries. The measured periods are listed in Table \ref{table: period and t_0} and their distribution is shown in Figure \ref{fig: basic properties}. We find that our sample has a fairly uniform period distribution with a slight excess of systems with periods of $4 - 5$ hours.

To determine the orbital ephemerides, we fitted each light curve with a model calculated with \texttt{ellc} \citep{ellc} for a detached eclipsing binary containing a hot sdB star and a cool MS companion. The primary goal was to determine the time of the primary eclipse, $t_0$, when the MS star moves in front of the sdB. We time the first eclipse after JD 2459000, which is near the midpoint of the light curve data for most targets, in order to minimize correlations between $t_0$ and $P_{\rm orb}$. We use the convention that phase 0 occurs at $t_0$, the time of the primary eclipse. 

We fixed the orbital period to the value found from the Lomb-Scargle periodogram, and model the reflection effect using the `heat' parameter (see \cite{ellc} for the details about the parameter). We report the best-fit $P_{\rm orb}$ and $t_0$ for each system in Table \ref{table: period and t_0}. We did not attempt to infer other physical parameters from the light curves at this stage, since our main goal is to measure the companion masses, and these are not directly constrained by the light curves.

\begin{table*}[htb!]
\centering
\begin{tabular}{ c c c c } 
 \hline
 Gaia ID & object name & period ($P_{\rm orb}$& eclipse time ($t_0$) \\ 
 & & [hours] & [HJD] \\
 \hline
  1375814952762454272 & J1533+3759 & 3.882492352179863 & 2459000.0486605708 \\
  2051078953817324672 & J1920+3722 & 4.054919786574039  & 2459000.1504712766 \\
 2995717462506292736 & J0557-1409 & 4.096633792122831 & 2459000.112196558\\
  2993468995592753920 & J0619-1417 & 4.224256397090405 & 2459000.0407816754\\
  4507223312777873280 & J1852+1445 & 4.570024569851073 & 2459000.083174447 \\
 4508520908289527808 & J1831+1345 & 4.741903317052129 & 2459000.2981252046\\
   2003241230122936064 & J2240+5437& 5.661092569611973 & 2459000.0809536236 \\
 4657996005080302720 & J0531-6953 & 6.276969603308999 & 2459000.14753494 \\
   2969438206889996160 & J0519-1916 & 6.59016169200534 & 2459000.13163848\\
 4467130720760209152 & J1630+1801 & 7.4232094820660715 & 2459000.1136060352\\
 6652952415078798208 & J1802-5532 & 8.663582790263654 & 2459000.0554469298 \\
 901929564359845888 & J0808+3202 & 8.873264194045575 & 2459000.231370364 \\
 4647004122914240640 & J0241-6855 & 11.05915838812156 & 2459000.281179102\\
 2943004023214007424 & J0612-1740 & 11.720487447561574 & 2459000.4836920444\\
 \hline
\end{tabular}
\caption{Orbital ephemerides for the sdB + MS binaries in our sample. The period was determined from a Lomb-Scargle periodogram, and the eclipse time was determined by fitting the light curve with a detached binary model calculated with \texttt{ellc} (see Section \ref{sec: period and eclipse time}).}
\label{table: period and t_0}
\end{table*}

\subsubsection{Observable parameters}
\label{sec:observables}
\begin{figure*}[ht!]
\includegraphics[width = \textwidth]{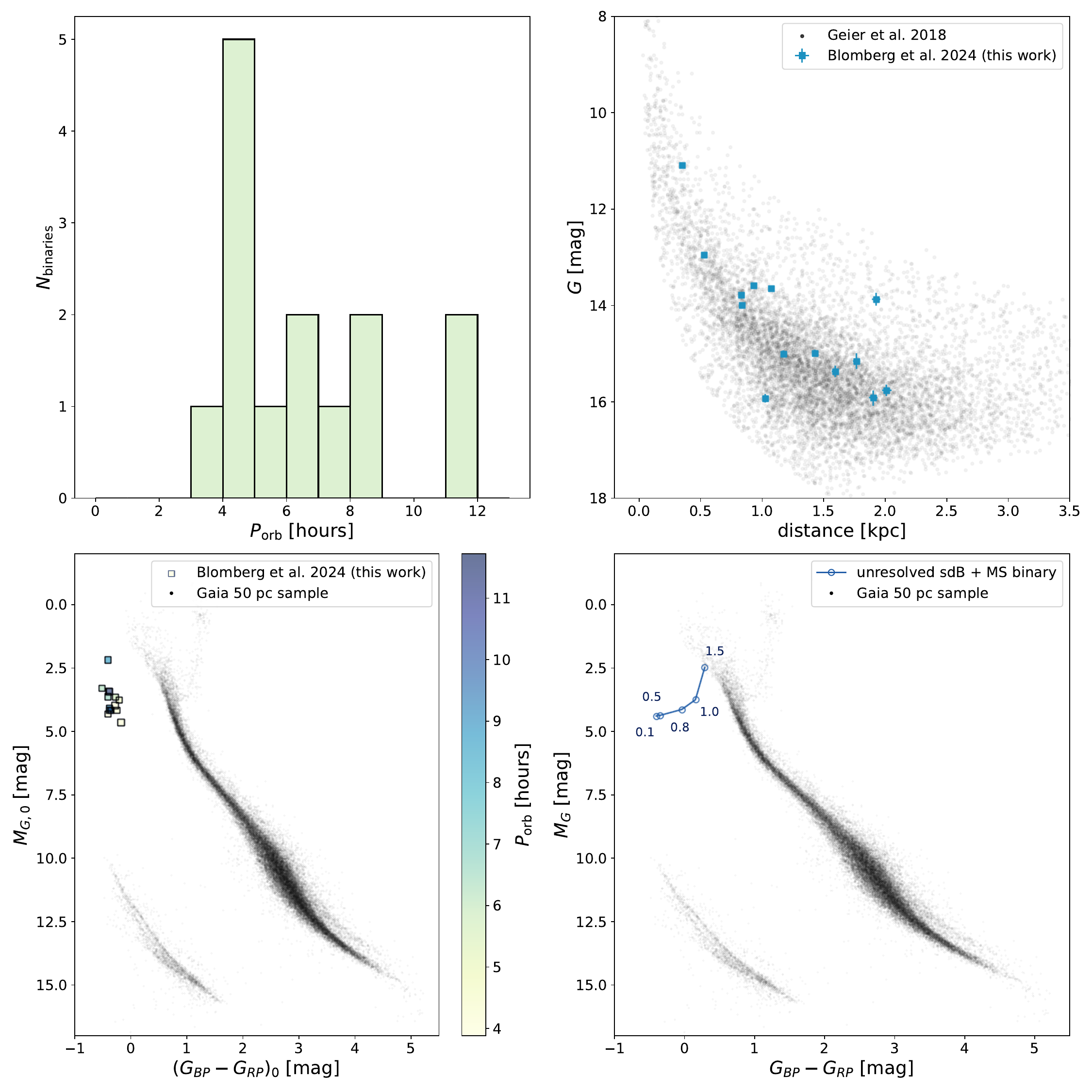}
\caption{Top left: orbital period distribution of objects in our sample. Top right: distance (i.e., $1/\varpi$) and $G-$band apparent magnitude. Our sample is compared to objects in the sdB candidate catalog from \citet{Geier2020} with $\varpi/\sigma_{\varpi} > 5$. The binaries in our sample are all within $\sim$2 kpc of the Sun and have apparent magnitudes ranging from $\sim16-11$ mag. Bottom left: objects in our sample on an extinction-corrected Gaia CMD, compared to the 50 pc sample. They all fall in the sdB clump, with little light contributions from the MS stars in the optical. Bottom right: predicted photometry for unresolved sdB + MS binaries with a range of companion masses. Hollow points shows predictions for a typical sdB paired with a MS star of mass 0.1, 0.5, 0.8, 1.0, and 1.5 $M_\odot$. Only companions with masses above 0.8 $M_\odot$ are predicted to appreciably move the unresolved source away from the sdB clump.}
\label{fig: basic properties}
\end{figure*}

We obtained other basic parameters of our targets from Gaia DR3 \citep{Gaia_dr3}, including their parallax, $G$-band apparent magnitude, and $G_{\rm BP} - G_{\rm RP}$ color. These values are reported in Table \ref{table:basic_properties}. We obtained extinction estimates for our targets from the 3D dust maps of \citet{extinction_north} and \citet{extinction}. We show the sources in an extinction-corrected color-magnitude diagram (CMD) in the bottom panel of Figure \ref{fig: basic properties}. For reference, we also show the Gaia 50 pc sample. The objects in our sample fall in the sdB clump, blueward of the MS and above the white dwarf cooling track.

The \cite{Schaffenroth_2022} sample of sdB candidates from which our sample is drawn were selected on the basis of their position in the sdB clump of the CMD. Our sample is thus biased against sdB + MS binaries that do not fall in the clump. To assess the effects of this bias, we show predicted photometry of unresolved sdB + MS binaries with a range of MS star masses in the bottom left panel of Figure \ref{fig: basic properties}. We assume absolute magnitudes of $G$, $G_{\rm BP}$, $G_{\rm RP}$ = 4.4, 4.2, 4.6 mag for the sdB, and we take the predicted magnitudes for the MS star from \cite{Pecaut_2013}. We find that sdB + MS binaries are predicted to fall redward of the sdB clump only for MS star masses $M_{\rm MS}\gtrsim 0.8\,M_{\odot}$. This means that our sample is biased against MS companions more massive than $\sim 0.8 M_{\odot}$ \citep[which are expected to form in wide orbits through stable mass transfer; e.g.][]{Han_2002}, but not against companions with lower masses. Given their CMD position, we expect MS star masses below $0.8 M_\odot$ for the systems in our sample.

\begin{center}
\begin{table*}
\begin{tabular}{ c c c c c c c c }
 \hline
  object name & RA & Dec & $G$ & $M_{\rm G}$ & $G_{\rm BP}-G_{\rm RP}$ & $\varpi$  & $E(B-R)$ \\ & [deg] & [deg] & [mag] & [mag] & [mag] & [mas] & [mag] \\
 \hline
  J1533+3759 &233.45601 & 37.99106& 12.94 & 4.34 $\pm$ 0.04 & -0.384 & 1.904 $\pm$ 0.035 & 0.02\\
  J1920+3722 &290.24905 & 37.37222& 15.76 & 4.25 $\pm$ 0.12 & -0.141 & 0.498 $\pm$ 0.029 & 0.11\\
  J0557-1409 &89.37149 & -14.16629& 15.92 & 5.87 $\pm$ 0.09 & 0.440  & 0.977 $\pm$ 0.040 & 0.47\\
  J0619-1417 & 94.76123& -14.28690& 15.91 & 4.52 $\pm$ 0.16 & -0.075 & 0.527 $\pm$ 0.038 & 0.14\\
  J1852+1445 &283.03169& 14.76307 & 14.99 & 4.65 $\pm$ 0.07 & 0.239  & 0.852 $\pm$ 0.025 & 0.34\\
 J1831+1345 & 277.81563 & 13.75534 & 15.15 & 3.92 $\pm$ 0.17 &-0.137 & 0.567 $\pm$ 0.045 & 0.19\\
   J2240+5437 & 340.21327 & 54.63084 & 14.97 & 4.21 $\pm$ 0.07 & 0.011 & 0.701 $\pm$ 0.023 & 0.22\\
 J0531-6953 & 82.91801 & -69.88371 &11.10 & 3.36 $\pm$ 0.03 & -0.478 & 2.837 $\pm$ 0.044 & 0.03\\
   J0519-1916 & 79.94864 & -19.28166& 13.59 & 3.75 $\pm$ 0.07 & -0.349 & 1.077 $\pm$ 0.032 & 0.05\\
 J1630+1801 & 247.68937 & 18.02232 & 15.37 & 4.36 $\pm$ 0.12 & -0.271 & 0.628 $\pm$ 0.034 & 0.08\\
 J1802-5532 & 270.70734 & -55.54981 & 13.83 & 2.45 $\pm$ 0.13 & -0.275 & 0.53 $\pm$ 0.03 &  0.10 \\
 J0808+3202 & 122.11083 & 32.04179 & 13.78 & 4.19 $\pm$ 0.06 & -0.323 & 1.205 $\pm$ 0.036 & 0.05\\
 J0241-6855 & 40.32553 &-68.92373 &14.64 & 3.50 $\pm$ 0.06 & -0.354 & 0.933 $\pm$ 0.027 & 0.03\\
 J0612-1740 & 93.19719 & -17.67511 &13.99 & 4.39 $\pm$ 0.05 & -0.241 & 1.198 $\pm$ 0.026& 0.09\\
 \hline
\end{tabular}
\caption{Basic properties of sdB + MS binaries in our sample. The parallax, $M_G$, and $G_\textrm{BP}-G_\textrm{RP}$ values are taken from Gaia DR3 \citep{Gaia_dr3}; the reddening values are taken from \citet{extinction_north} and \citet{extinction}.}
\label{table:basic_properties}
\end{table*}
\end{center}

\section{Follow-up Observations} \label{sec: follow-up observations}
To constrain the MS star masses, we obtained multi-epoch follow-up spectroscopy of all objects in our sample using several different instruments. We measured RVs of each target in at least three epochs obtained across at least two nights.

We used four spectrographs at three different observatories: the Low Resolution Imaging Spectrometer \citep[LRIS;][]{LRIS} and Echellette Spectrograph and Imager \citep[ESI;][]{ESI} on the 10m Keck I and Keck II telescopes, the Double Spectrograph \citep[DBSP;][]{DBSP} on the 5m Hale telescope at Palomar observatory, and the Fiber-fed Extended Range Optical Spectrograph \citep[FEROS;][]{FEROS} on the ESO/MGP 2.2m telescope at La Silla Observatory. We used LRIS, ESI, and DBSP for targets in the north, and FEROS for targets in the south. Below, we discuss the settings and data reduction process for each instrument. 

\subsection{Low Resolution Imaging Spectrometer (LRIS)}
For LRIS, we observed on the blue and red arms simultaneously using a 1.0 arcsecond slit. We used exposure times of 300 and 600 seconds (depending on the magnitude of the target), and obtained spectra covering wavelengths of $\lambda = 320 - 1000$ nm with a resolution of $R \sim 2000$. The data were reduced using LPipe \citep{LPipe}. 

\subsection{Echellette Spectrograph and Imager (ESI)}
For ESI, we used the 0.3 arcsecond slit with $2 \times 1$ binning and 300 second exposures. We obtained spectra covering $\lambda = 390 - 1100$ nm at a resolution of $R \sim 11,000$. We reduced the data using the MAuna Kea Echelle Extraction (MAKEE) pipeline.

\subsection{Fiber-fed Extended Range Optical Spectrograph (FEROS)}
From FEROS, we obtained spectra covering wavelengths $\lambda = 360-920$ nm at a resolution of $R \sim 50,000$. We used exposure times of $1800-2400$ seconds. The data were reduced using the CERES pipeline \citep{CERES}.

\subsection{Double Spectrograph (DBSP)}
For DBSP, we used a 1.0 or 1.5 arcsecond slit (depending on seeing) and exposure times of 300 to 600 seconds (depending on the magnitude of the target). Using both the blue and red arms, we obtained spectra covering the wavelength range of $\lambda = 350 - 800$ nm at a resolution of $R \sim 1,500$. To reduce the data, we used the \texttt{pypeit} reduction pipeline \citep{pypeit}. We also applied an empirical correction for instrumental flexure to the wavelength solution using telluric absorption lines, as described by \cite{Nagarajan_2023}.

All our observations are listed in Table \ref{table: radial velocities} in Appendix \ref{adx: RV measurements}.

\section{Analysis}\label{sec: analysis}
\subsection{Radial velocity measurements}
We normalized the spectra using a running median calculated in a 101 $\rm \AA$ window. We normalized the template spectra using the same method. The sdB stars dominate the observed spectra, and we do not detect absorption features from any of the MS companions.

We measured the RV of the sdB star in each spectrum by cross-correlating a synthetic template with the data after applying a barycentric correction. For spectra from DBSP, ESI, and LRIS, we used templates from \cite{spectra_template}, who calculated a grid of non-local thermodynamic equilibrium (NLTE) synthetic spectra covering $\lambda = 320 - 720$ nm using Tlusty and Synspec \citep{TLUSTY}. Their grid provides models for a range of effective temperatures, surface gravities, and surface He abundances. Our primary goal was to measure reliable RVs. This requires a template with similar spectral lines to the observed spectra, but not necessarily one whose parameters match the true parameters of each observed sdB. Because multiple combinations of atmospheric parameters and abundances can produce similar spectra, we fixed the temperature and surface gravity to typical values for sdBs: $T_{\rm eff}=35,000$ K and $\log(g / [\textrm{cm/s}^2]) = 5.0,$ and chose the model for each observed sdB with He abundance that best matched the observed spectra. We then used the same template for all observations of a given system.

\begin{figure*}[ht!] 
\includegraphics[width = \textwidth]{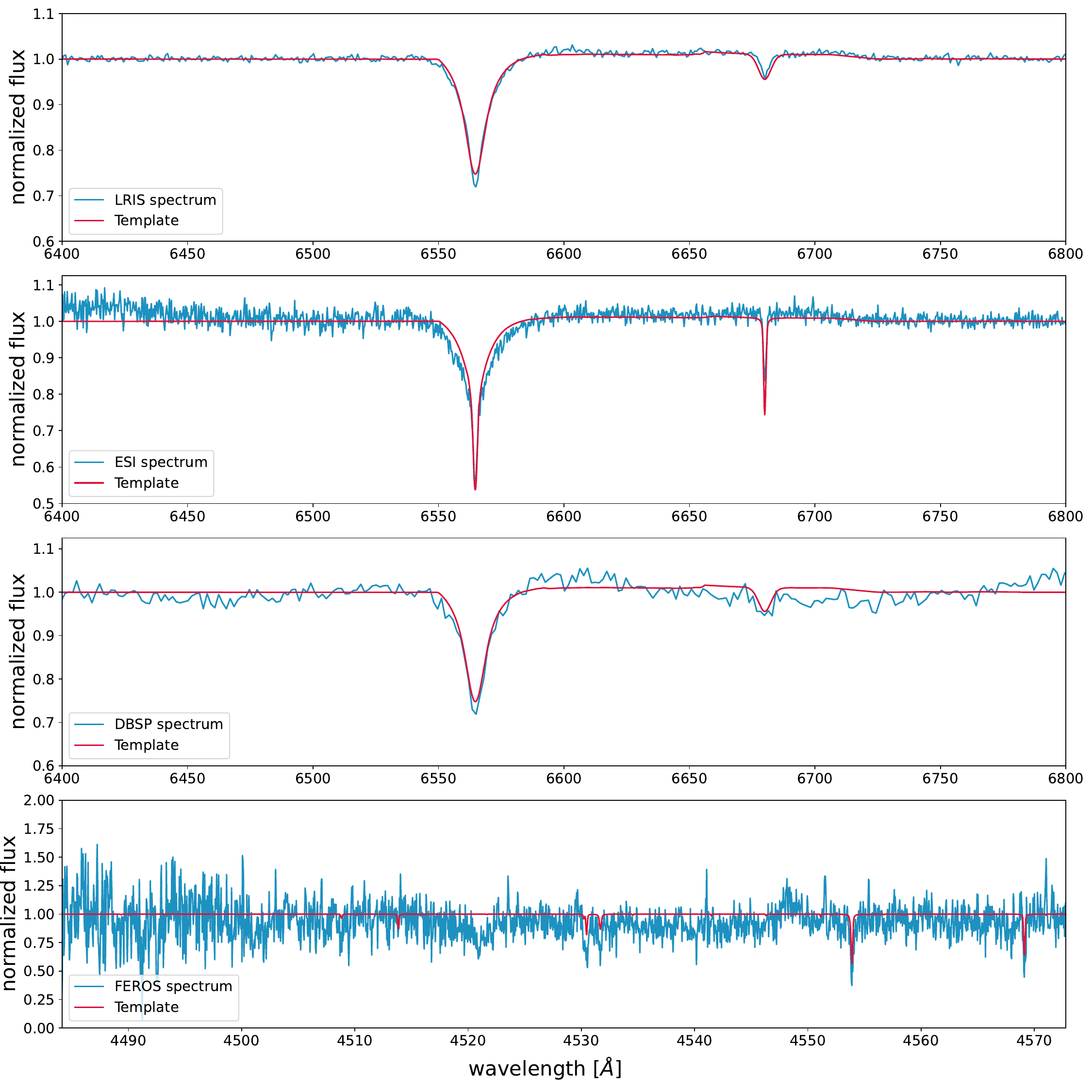}
\caption{Example spectra from LRIS, ESI, DBSP, and FEROS. We overplot a model spectrum with the inferred RV in each panel. The top three panels show spectra of J0519-1916, and the best-fit synthetic spectrum from \cite{spectra_template}. The H$\alpha$ and He I lines provide the primary RV constraint. The bottom plot show a FEROS spectrum for J1802-5532 and corresponding model spectrum; here the RV constraint comes from narrow metal lines.}
\label{fig: spectra fits}
\end{figure*}

We cross-correlated the template and observed spectra over wavelengths $\lambda = 640 - 680$ nm, which contain the H$\alpha$ absorption line ($\lambda$ = 656.46 nm) and a strong He I absorption line ($\lambda$ = 667.82 nm). For spectra taken with FEROS, which have higher resolution and therefore contain several resolved narrow metal lines, we instead used the wavelength range of $\lambda = 450 - 550$ nm, which contains several narrow metal lines. Since these metal lines are not present in the \cite{spectra_template} models, we instead used a bespoke template generated with Tlusty and Synspec \citep{TLUSTY}, assuming solar metallicity. We convolved each of the spectral templates with a Gaussian kernel of appropriate FWHM to account for instrumental broadening for each instrument. The formal RV uncertainties are small ($\lesssim 1 \textrm{km s}^{-1}$). However, we expect the true uncertainties to be dominated by systematics, such as flexure and zerpoint offsets between different instruments. Therefore, we adopted conservative uncertainties of 10 km/s for DBSP and LRIS, and 5 km/s for ESI and FEROS, based on the RV stability of standard stars observed with the same setup. The RV measurements for all of the spectra are listed in Table \ref{table: radial velocities} in Appendix \ref{adx: RV measurements}. Some of our fits, representative of our observed population, are shown in Figure \ref{fig: spectra fits} for reference.

\subsection{RV curves}
\begin{figure*}[ht!]
\includegraphics[width = 0.925 \textwidth]{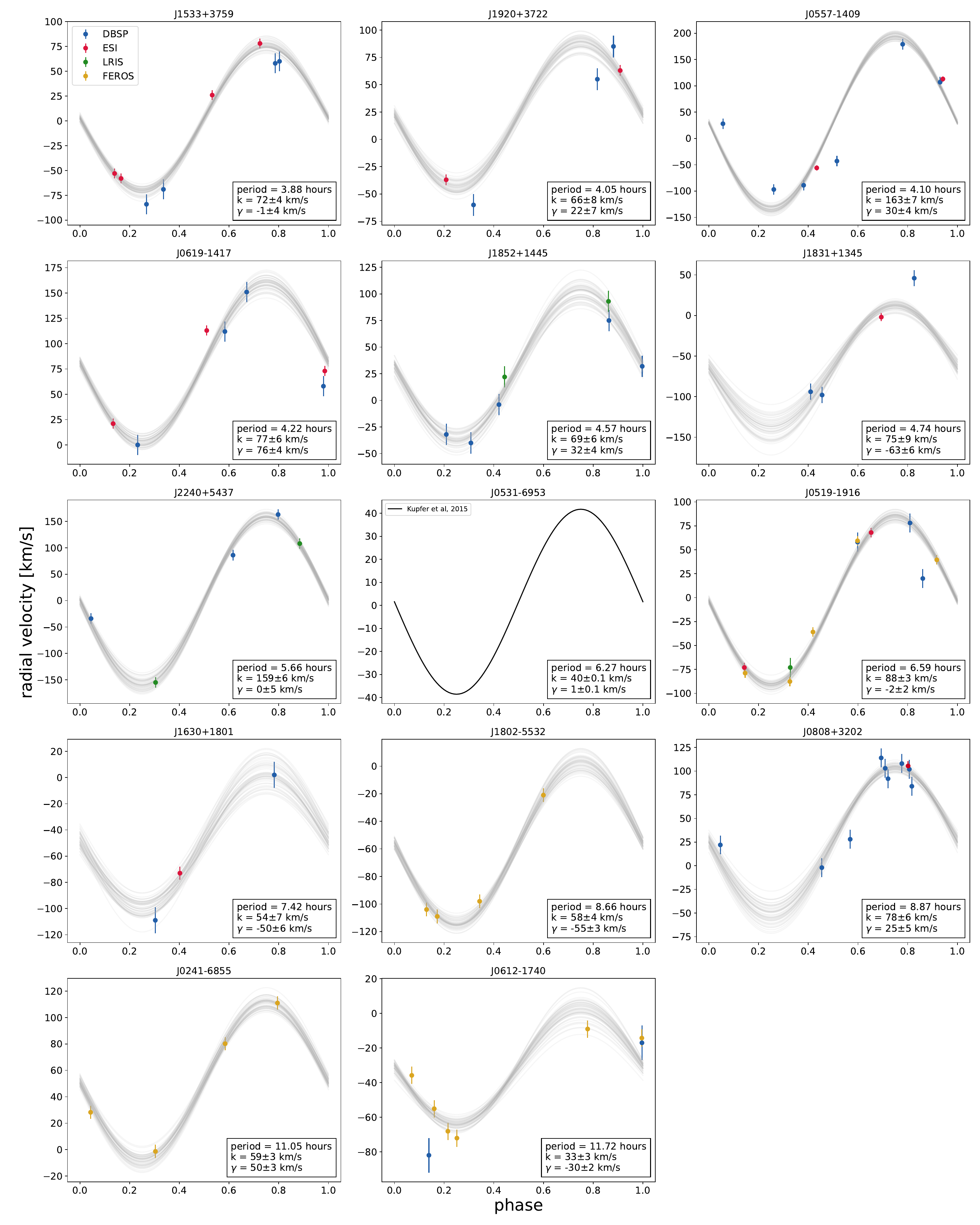}
\caption{Measured RVs as a function of orbital phase for each target. The colors of data points correspond to which instrument the measurement was made with. We fit the observed RVs with a sinusoidal model (Equation \ref{eq: RV}), fixing the orbital period and eclipse time to the value measured from the light curve. This allows us to constrain the RV curves robustly even with only a few data points. Gray lines show random samples from the posterior. The corresponding constraints are listed in Table \ref{table: radial velocity parameters}. The RV curve for J0531-6953 is taken from a previous study by \cite{Kupfer_2015}.}
\label{fig: radial velocity fits}
\end{figure*}

Using the ephemerides obtained from the light curves, we fit the observed RVs to constrain the RV semi-amplitudes of the sdB stars and the binaries' center-of-mass velocities. For each target, we fitted the measured RVs with a sinusoid:
\begin{equation}\label{eq: RV}
    {\rm RV} = - K \sin\left(\frac{2\pi (t-t_0)}{P_{\rm orb}}\right) + \gamma,
\end{equation}
where $K$ is the semi-amplitude of the velocity variation of the sdB, $t_0$ is the time of the primary eclipse (Table \ref{table: period and t_0}), $P_{\rm orb}$ is the orbital period, and $\gamma$ is the center of mass velocity. Equation \ref{eq: RV} assumes a circular orbit, which is expected for our systems since the periods are short enough to be circularized by tides. We fit the data using \texttt{emcee} \citep{EMCEE} to sample from the posterior, adopting broad flat priors on the free parameters $K$ and $\gamma$. The results of the fit are shown in Figure \ref{fig: radial velocity fits}, and the inferred parameters are listed in Table \ref{table: radial velocity parameters}. For J0531-6953, we show the best-fit RV curve measured by \citet{Kupfer_2015}.

\subsection{Companion masses} \label{sec: companion mass measurement}
\begin{figure*}[ht!]
\includegraphics[width = \textwidth]{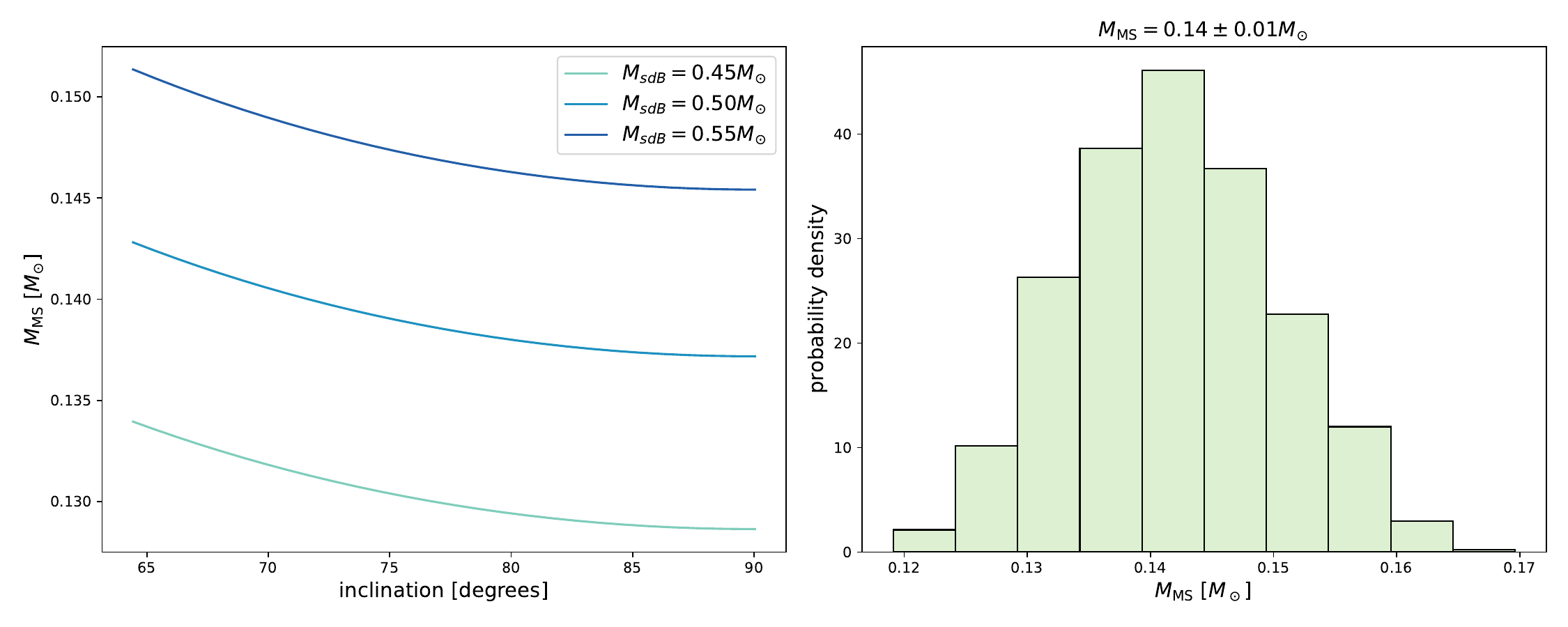}
\caption{Companion mass constraints for J1533+3759, a typical object in our sample. Left: companion mass implied by the sdB RV mass function (Equation \ref{eq: mass function_2}) for various sdB masses $M_\textrm{sdB} = (0.45 - 0.55) M_{\odot}$ and inclination angles $i$ = 64.4 - 90.0 degrees. Right: the companion mass distribution produced from 100 samples of sdB mass from $\mathcal(U)(0.45, 0.55) M_{\odot}$, 100 samples of inclination from a $\sin(i)$ inclination distribution for eclipsing systems ranging from $64.43 - 90.0$ degrees, and 50 samples of the RV semi-amplitude $K$ from the posterior of the MCMC fitting. The median and standard deviation of the inferred $M_{\rm MS}$ values are reported as the best-fit companion mass and its uncertainty.}
\label{fig: mass function}
\end{figure*}

Our constraints on the sdB stars' RV semi-amplitudes allow us to constrain the companion masses, subject to assumptions about the inclination and the sdB mass. We first compute the RV mass function,
\begin{equation}\label{eq: mass function_1}
    f_m = \frac{P_{\rm orb}K^3}{2 \pi G}.
\end{equation}
The mass function is related to the component masses and orbital inclination:
\begin{equation}\label{eq: mass function_2}
    f_m = \frac{{M_{\rm MS}^2} \sin^3i}{\left(M_{\rm sdB}+M_{\rm MS}\right)^2}.
\end{equation}
$P_{\rm orb}$ and $K$ are constrained by the light curves and RV measurements, respectively. Further, since we limited our study to eclipsing systems, we expect the inclinations to be close to 90 degrees. For a given choice of sdB mass and inclination, we can then numerically solve Equation \ref{eq: mass function_2} for the mass of the MS companion. Figure \ref{fig: mass function} illustrates how our constraints on $P_{\rm orb}$ and $K$ then translate to constraints on $M_{\rm MS}$. Higher assumed sdB masses and lower assumed inclinations translate to higher MS star masses, but the dependence on both of these quantities are weak within their plausible ranges for eclipsing systems.

To determine the uncertainties on the companion mass, we produced a distribution of masses using a Monte Carlo simulation. For each target, we randomly sampled 50 RV semi-amplitudes from the MCMC posterior distribution produced when fitting the RV curves. Further, we sampled 100 sdB masses from a uniform distribution, $\mathcal{U}(0.45, 0.55)\,M_{\odot}$ \citep{han_2003}. Then for each sdB mass in our sample, we sampled 100 inclination angles from a $\sin(i)$ inclination distribution between $i_{\rm min}$ and 90 degrees, where $i_{\rm min}$ is the minimum possible inclination angle for a system with a given period to eclipse. From the geometry of the system, we can solve for the minimum inclination angle, $i_{\rm min}$, using the relation:
\begin{equation}
    \cos{(i_{\rm min})} = \frac{R_{\rm MS} + R_{\rm sdB}}{a}.
\end{equation}
Using Kepler's laws, we can rewrite this as:
\begin{equation}\label{eq: min inclination angle}
    \cos{i_{\rm min}} = \left (R_{\rm MS} + R_{\rm sdB}\right ) \left(\frac{4\pi^2}{GP_{\rm orb}(M_{\rm MS} + M_{\rm sdB})}\right)^{1/3}.
\end{equation}
We took a typical sdB mass ($M_{\rm sdB} = 0.5 M_\odot$) and a radius $R_{\rm sdB} = 0.21R_{\odot}$, corresponding to a surface gravity $\log(g [\textrm{cm/s}^2]) = 5.5$. For the MS star, we took a mass $M_{\rm MS} = 0.3 M_\odot$ and radius $R_{\rm MS} = 0.29 R_\odot$, as predicted by MIST isochrones \citep{MIST}. This value is conservative, since we will show that most MS stars in the sample have masses lower than 0.3 $M_\odot$, and $i_{\rm min}$ corresponds to very shallow grazing eclipses. For each system, we calculated $i_{\rm min}$ by solving Equation \ref{eq: min inclination angle} using $P_{\rm orb}$ determined from the light curves. For each $K$, we numerically solved Equation \ref{eq: mass function_2} for each combination of the sampled sdB mass, inclination, and mass function. This procedure produces an approximately Gaussian distribution of dynamically-implied MS star masses (right panel of Figure \ref{fig: mass function}). We report the median and standard deviation of this distribution as the best-fit MS star mass and corresponding uncertainty in Table \ref{table: radial velocity parameters}.

Our assumption of  $M_{\rm sdB} \sim \mathcal{U}(0.45, 0.55)\,M_{\odot}$ is reasonable for the canonical sdB formation channel, but some sdBs can have higher or lower masses \citep[e.g.][]{Han_2002}, and we cannot rule out this possibility for any given system. To assess the impacts of the assumed sdB mass on our results, we explore more extreme masses of $M_{\rm sdB} = 0.35 M_\odot$ and $M_{\rm sdB} = 0.60 M_\odot$ in Appendix~\ref{appendix: companion mass}. We find that our results are only weakly sensitive to the assumed sdB mass even over this rather broad range.

\begin{table}[ht!]
\centering
 \begin{tabular}{c c c c} 
 \hline
 object name & $K$ [km s$^{-1}$] & $\gamma$ [km s$^{-1}$] &  $M_{\rm MS}$ [M$_\odot$]\\ [0.5ex] 
 \hline  \\[0.1ex] 
  J1533+3759 & $72.39 \pm 4.14$ & $-1.87 \pm 3.53$ & $0.14\pm0.01$
  \\[1.0ex] 
  J1920+3722 & $66.14 \pm 3.80$ & $22.78\pm 3.05$ & $0.12\pm0.01$  
  \\[1.0ex]     
  J0557-1409 & $163.88\pm 5.09$ & $30.54\pm 2.70$ & \textbf{$0.41\pm0.02$}
  \\[1.0ex] 
  J0619-1417 & $77.15\pm 6.16 $ & $76.28\pm3.79$ & $0.16\pm0.01$
  \\[1.0ex] 
  J1852+1445 & $69.33 \pm 5.64$ & $32.69 \pm 3.89$ & $0.14\pm0.01$
  \\[1.0ex] 
  J1831+1345 & $75.54 \pm 6.25$ & $-63.30 \pm 5.08$ & $0.16\pm0.02$ \\[1.0ex]
  J2240+5437 & $159.65 \pm 6.27$ & $-0.41 \pm 4.66$ & $0.46\pm0.03$ \\[1.0ex] 
  J0531-6953* & $40.15\pm0.11$ & $1.59\pm0.08$ & $0.09\pm0.01$ \\[1.0ex] 
  J0519-1916 & $88.84 \pm 2.83$ & $-2.85\pm 1.98$ & $0.22\pm0.01$\\[1.0ex] 
  J1630+1801 & $54.95 \pm 6.89$ & $-50.09 \pm 5.92$ & $0.12\pm0.02$\\[1.0ex]
  J1802-5532 & $58.62 \pm 4.10$ & $-55.74 \pm 3.12$ & $0.15\pm0.02$\\[1.0ex]
  J0808+3202 & $78.82 \pm 6.16$ & $-25.77 \pm 5.23$ & $0.21\pm0.02$\\[1.0ex]
  J0241-6855 & $59.82 \pm 3.45$ & $50.70 \pm 2.52$ & $0.17\pm0.01$ \\[1.0ex]
  J0612-1740 & $33.54 \pm 2.85$ & $-30.49 \pm 2.23$ & $0.09\pm0.01$\\[1.0ex]
 \hline
 \end{tabular}
 \caption{sdB RV semi-amplitudes and center-of-mass RVs measured from our follow up. We calculate the companion mass and its uncertainty from Monte Carlo samples of the semi-amplitude, orbital period, inclination, and sdB mass (see Figure \ref{fig: mass function}). *Parameters for J0531-6953 were taken from \cite{Kupfer_2015} and propagated to constraints on $M_{\rm MS}$ using the same procedure as for the objects we followed-up.}
 \label{table: radial velocity parameters}
\end{table}

\subsection{Population properties}
\begin{figure*}[ht!]
\includegraphics[width = \textwidth]{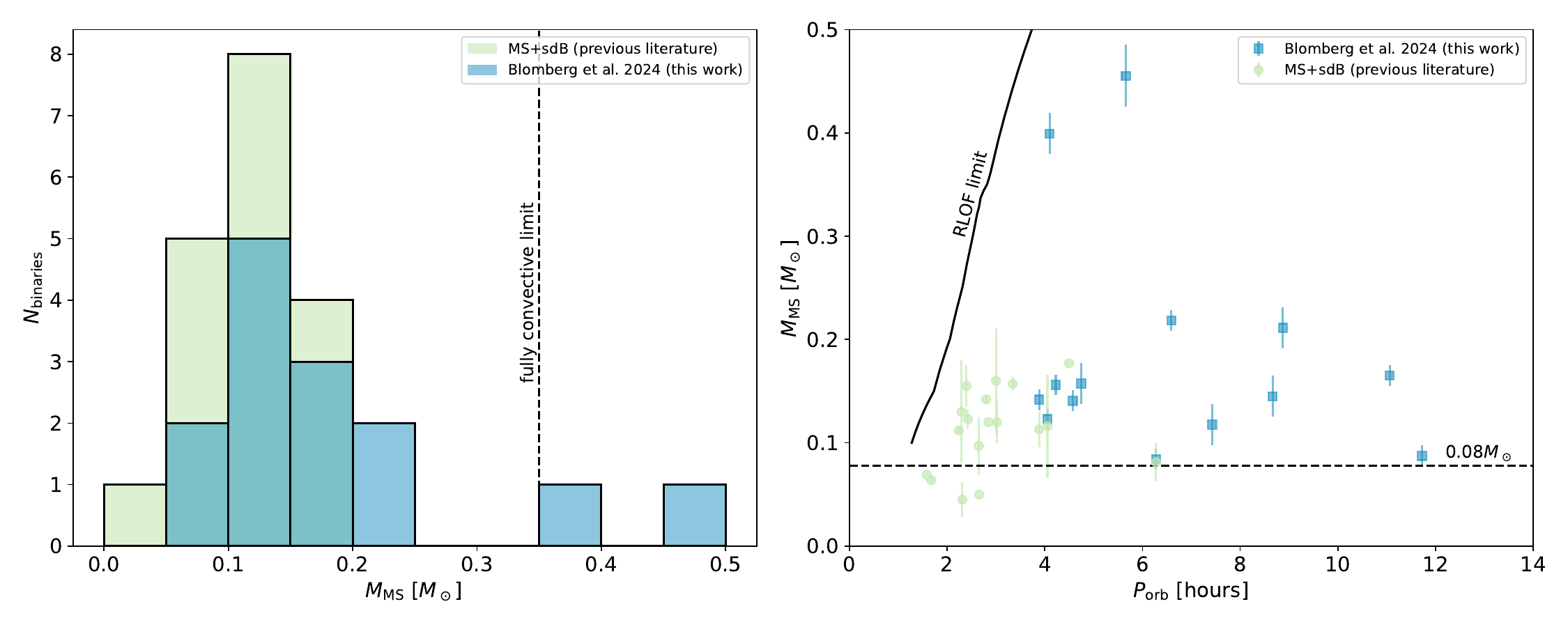}
\caption{Left panel: mass distribution of the MS companions to sdB stars. We compare our sample (blue) to a literature compilation by \citet[][green]{Schaffenroth_2018}. Dashed lined marks the fully convective limit. The mass distribution is peaked near $0.15 M_\odot$ in both samples, with a drop-off above $0.2 M_\odot$ (somewhat lower than the fully convective limit), though our sample also contains two systems with $M_{\rm MS}\sim0.4M_\odot$. Right panel: companion mass vs. orbital period. Dashed line marks the hydrogen burning limit and the solid line marks the Roche lobe overflow limit. Compared to the previous literature, our sample probes longer orbital periods, which could have accommodated more massive MS companions than the short-period systems studied in previous work. However, most of the objects in our sample have low MS star masses, similar to previous literature samples.}
\label{fig: population analysis}
\end{figure*}

Figure \ref{fig: population analysis} shows the inferred companion masses and orbital periods of binaries in our sample. We compare to other HW Vir binaries with companion mass constraints from the literature compilation by \cite{Schaffenroth_2018}.

The companion mass distribution for both samples of HW Vir binaries peaks at $0.1 - 0.15 M_\odot$ and falls off above $0.2\,M_{\odot}$. This is reminiscent of the lack of higher-mass MS stars in WD + MS binary samples that has been interpreted as evidence for DMB \citep[e.g.][]{Schreiber_2010}. Unlike the previous literature, our sample does contain two binaries with companion masses near $0.4\,M_{\odot}$,  above the fully convective limit. However, the bulk of the population has lower $M_{\rm MS}$.

The right panel of Figure \ref{fig: population analysis} shows the companion masses and orbital periods of sdB + MS binaries from both samples. The black line marks the Roche lobe overflow limit. Compared to the previous literature, our sample includes a wider range of orbital periods and companion masses. Crucially, we focused on long orbital periods ($P_{\rm orb} > 3.8$ hours) because these orbits are wide enough to accommodate a wide range of MS star masses. In contrast, most of the previously studied systems have orbital periods of $1 - 3$ hours, which can only accommodate low-mass companions ($M_{\rm MS} \lesssim 0.3\,M_{\odot}$), since higher-mass MS stars would overflow their Roche lobes (black line). Below the fully convective limit, the mass distribution of our sample is similar to that found at shorter periods in the literature. There is no significant correlation between companion mass and orbital period in our sample.

\subsection{Selection effects and biases}
Our sample was selected to (a) be brighter than $G = 16$ mag, (b) have $P_{\rm orb} = 3.8 - 12$ hours, and (c) be eclipsing. We can model the effects of each cut. 

Given that typical sdBs have $M_{G,0}\approx 4$ \citep{Heber_2016}(a) translates to a distance limit of $\approx 2.5$ kpc, or closer for systems with significant foreground extinction. We do not expect $M_{G,0}$ to be significantly correlated with companion mass, since the MS companions contribute $\lesssim 1\%$ of the optical light for $M_{\rm MS} \lesssim 0.5\,M_\odot$.

\begin{figure}[htp]
    \centering
    \includegraphics[width= 0.5 \textwidth]{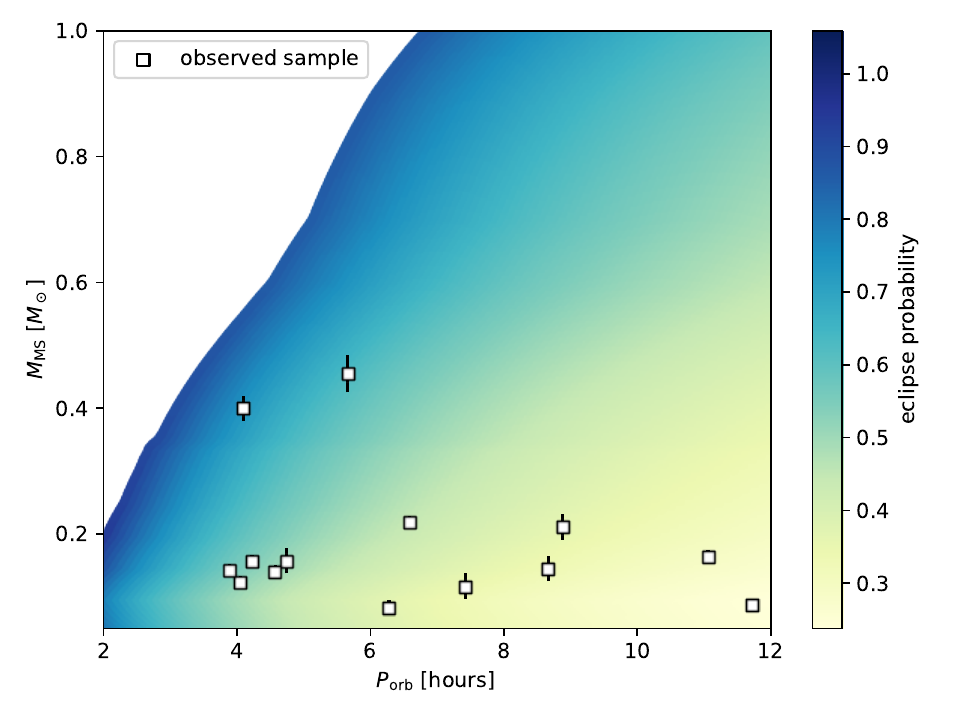}
    \caption{Eclipse probability for sdB + MS binaries, assuming randomly oriented orbits, $M_{\rm sdB} = 0.5\,M_{\odot}$, and $\log(g[\textrm{cm/s}^2]) = 5.5$. The white region in the upper left corresponds to binaries in which the MS star would overflow its Roche lobe. Our targets are shown with white squares. They have periods ranging from 3.8 to 12 hours -- long enough to accommodate MS star mass up to 0.45 $M_\odot$ at all periods, and up to 0.7 $M_\odot$ for $P_{\rm orb} \gtrsim 6$ hr. At fixed $P_{\rm orb}$, the eclipse probability is higher for higher-mass MS stars, which have larger radii. Our selection of eclipsing systems thus leads to a bias against high-mass companions. Nevertheless, low-mass companions dominate the observed sample, suggesting that they dominate the full population.}
    \label{fig: eclipse probability}
\end{figure}

The cut on orbital period, (b), corresponds to a range of allowed companion masses, since only low-mass MS stars are sufficiently dense to fit in the shortest-period orbits without overflowing their Roche lobes (Figure \ref{fig: eclipse probability}). However, this does not lead to a significant bias in our sample, because MS stars with masses $\lesssim 0.5 M_\odot$ could fit inside the full range of orbital periods represented in our sample.

Finally, the requirement that binaries be eclipsing leads to a quantifiable bias against low-mass MS companions. Appendix \ref{adx: selection function} shows how the eclipse probability varies with MS companion mass and orbital period. The mass of the sdB star has little effect on the eclipse probability, with less than 5\% variations in the eclipse probability for $M_{\rm sdB} = 0.3 - 0.6 M_\odot$ assuming $\log (g / [\textrm{cm/s}^2]) =5.5$ when the mass of the companion and the period are held constant. Figure \ref{fig: eclipse probability} shows the predicted eclipse probability as a function of orbital period and MS star mass, assuming a fixed sdB star mass of 0.5 $M_\odot$.

At a fixed period, the predicted eclipse probability rises monotonically with $M_{\rm MS}$,  since higher-mass MS stars have larger radii. Our targets span a wide range of periods but all have relatively low companion masses and thus fall in the regime of $20 - 30$\% eclipse probabilities. The lack of higher-mass MS stars in our sample cannot be a selection effect driven by our requirement that binaries be eclipsing, since eclipse probabilities favor high-mass companions. More massive MS companions also lead to higher-amplitude reflection effects that are more easily detectable. This means that the intrinsic drop-off in the companion mass distribution must be even stronger than what is found in our observed sample.

\subsection{Comparison to other PCEB populations}
\begin{figure*}[ht!]
    \centering
    \includegraphics[width=\linewidth]{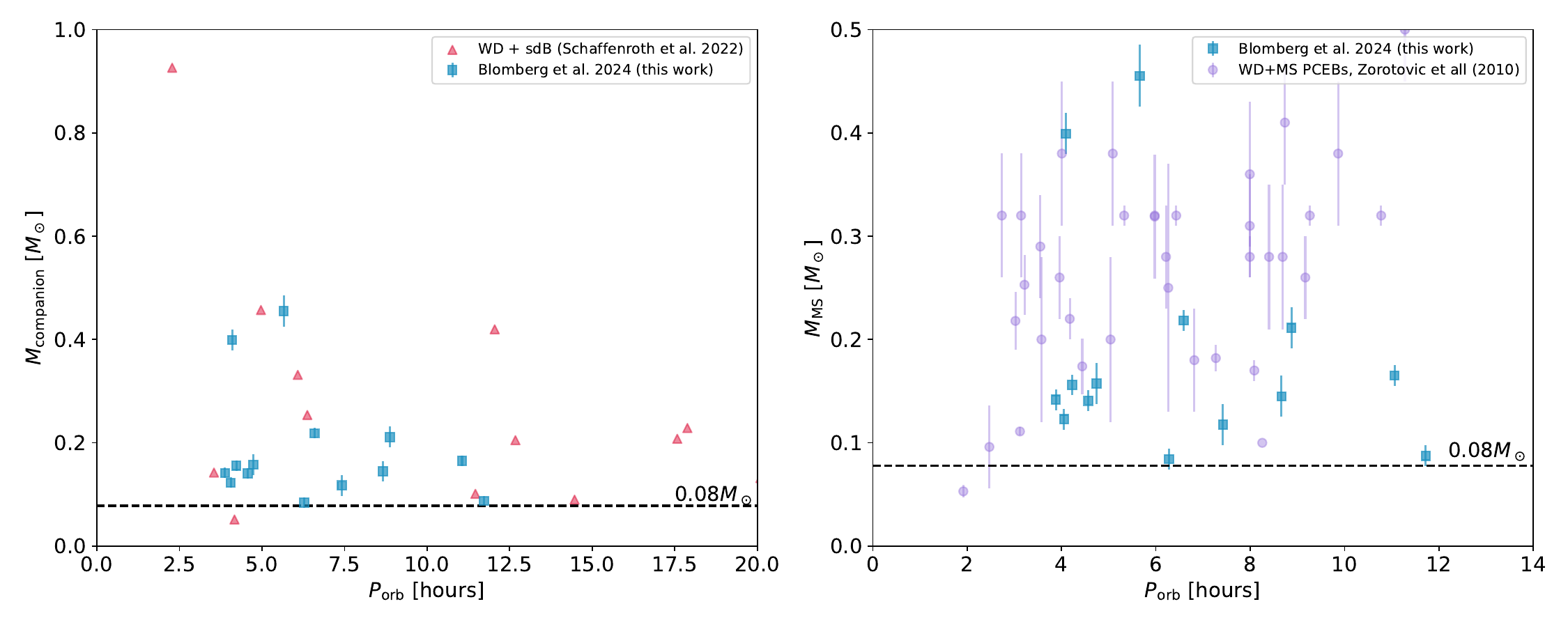}
    \caption{Left panel: observed sample compared to a population of sdB + WD systems from \cite{Schaffenroth_2022}. Dashed line marks the hydrogen burning limit. The WD masses from \cite{Schaffenroth_2022} are lower limits and we expect the true masses to be larger, near the typical WD mass of $\sim 0.5-0.6 M_\odot$. Right panel: observed sample compared to a population of WD + MS systems from \cite{zorotovic_2010}. The WD + MS systems are  distributed across a larger companion mass range ($\sim0.1 - 0.4M_\odot$), compared to the sdB + MS systems which tend to have lower companion masses.}
    \label{fig: population comparison}
\end{figure*}

Figured \ref{fig: population comparison} compares our observed sample to populations of sdB + WD (left panel) and WD + MS (right panel) binaries from \cite{Schaffenroth_2022} and \cite{zorotovic_2010}, respectively. 

The WD + sdB binary population represents a population of PCEBs that presumably have not undergone MB, since neither component of the binary is expected to experience MB. At $P_{\rm orb} > 3.8$ hours, the effects of gravitational wave angular momentum losses are also expected to be negligible during the $\sim100$ Myr lifetime of the sdB stars. Thus, we expect this population to be representative of the zero-age PCEBs that have not yet undergone any period evolution due to MB. Here, the measured WD masses from 
\cite{Schaffenroth_2022} are lower limits, since the inclinations are unknown, and we expect the true masses to be larger. This suggests that WD companions to sdB stars have significantly higher masses at shorter periods than do MS companions, similar to what \cite{Kupfer_2015} found in their analysis of sdB binaries. 

The WD + MS binary population represents a population of binaries that have presumably evolved under the same MB law as our sdB + MS sample. Given the longer observable lifetime of WDs, we expect this population to represent older binaries compared to our observed sample, and thus to have experienced more MB-driven orbital inspiral. Compared to our sdB + MS population, we find that the WD + MS population is uniformly distributed across a larger companion mass range ($\sim 0.1-0.4 M_\odot$) and drops off at a higher companion mass near $M_{\rm MS}\sim0.4M_\odot$. We note that selection biases for WD + MS binaries favor low-mass companions, since higher-mass companions outshine WDs. The excess of higher-mass MS companions to WDs relative to sdBs is thus unlikely to be driven by selection effects. This suggests that if the drop-off in the companion mass distribution toward higher masses is a result of DMB, then it operates even more strongly in sdB + MS binaries than in WD + MS binaries. We explore constraints on MB further below.

\section{Comparison to simulations} \label{sec: comparing observations to simulated populations}
To interpret the observed population and constrain formation models, we produced a synthetic population of close sdB + MS binaries using the Compact Object Synthesis and Monte Carlo Investigation Code \citep[COSMIC;][]{COSMIC} and evolved the simulated binaries under various MB prescriptions. We applied the selection function of our observed sample to the simulated population and then compared this simulated and mock-observed population to our observed sample. We tested two MB prescriptions: the RVJ MB model \citep{Verbunt_Zwaan_1981, RVJ_1983}, and the saturated MB model \citep{Kawaler_1988, Chaboyer_1995, Sills_2000}. We also experimented with re-scaling the strength of MB above and below the fully-convective boundary, following \cite{belloni_2023}.

\subsection{Producing an initial PCEB population with COSMIC}\label{sec: cosmic}
We used COSMIC to produce a population of zero-age PCEBs representative of the sdB + MS binaries in our sample. We assumed a \citet{Kroupa_2001} primary initial mass function, a uniform initial eccentricity distribution, and a lognormal initial period distribution following \citet{Raghavan_2010}. We simulate a single burst of star formation and evolve the resulting binary population for 13.7 Gyr. Since BSE \citep{BSE_2002} and COSMIC do not model the formation of sdB stars via the standard channels \citep[e.g.][]{Han_2002, han_2003}, these calculations do not directly form sdBs. Instead, when a primary overflows its Roche lobe near the tip of the first giant branch, it simply forms a He WD. Thus, we focus our analysis on PCEBs containing $0.4 - 0.5 M_\odot$ He WDs, many of which should have ignited core He burning near the tip of the first giant branch and formed sdB stars. We selected binaries that any point in their evolution contain $0.4 - 0.5 M_\odot$ He WDs (kstar\_1 = 10) and MS companions with masses $M_{\rm MS} < 0.7 M_\odot$ (kstar\_2 = 0). We explore the effects of selecting lower-mass sdBs, as might be expected for sdBs formed from progenitors that ignited helium non-degenerately, in  Appendix \ref{appendix: COSMIC larger mass range}. While this selection is not identical to the sdB binaries in our observed sample, the stellar properties at the onset of and emergence from the common envelope are sufficiently described to capture the companion mass and orbital period distribution of the sdB sample, even though the models do not capture the ignition of He in the core. We take this population produced by COSMIC to be the zero-age sdB + MS PCEB population.

\begin{figure}
    \centering
    \includegraphics[width=\linewidth]{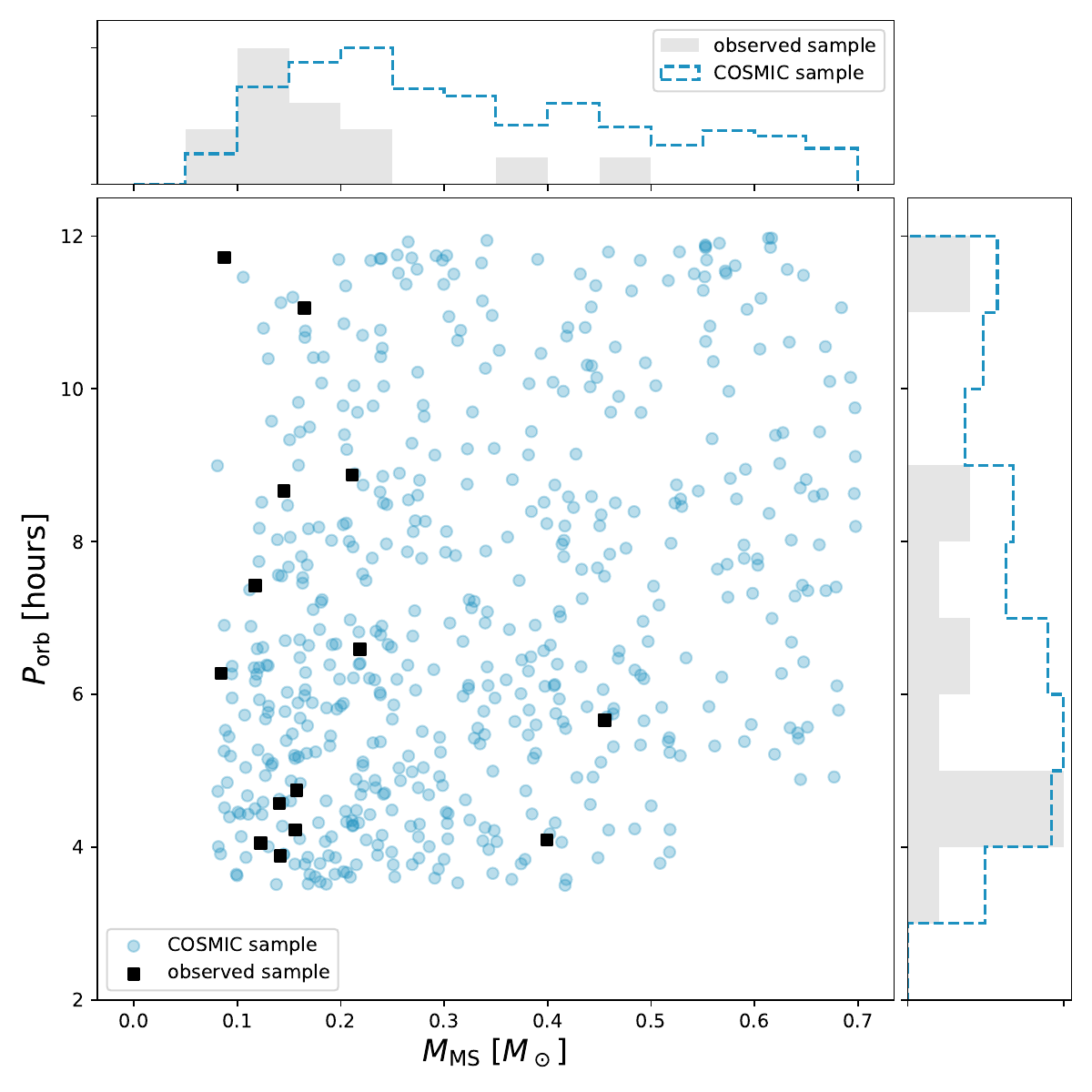}
    \caption{Comparison of the synthetic zero-age PCEB sdB + MS population produced using COSMIC with our observed sample. The zero-age PCEB population has a broad range of MS companion masses, unlike the observed population, which is dominated by companions with $M_{\rm MS} < 0.2 M_\odot$.}
    \label{fig: cosmic_population}
\end{figure}

Figure \ref{fig: cosmic_population} compares the synthetic population of sdB + MS binaries produced with COSMIC with our observed sample.
The zero-age PCEBs from COSMIC have a fairly uniform companion mass distribution, with only a gradual drop-off toward higher $M_{\rm MS}$, and significantly more high-mass companions than the observed sample. This suggests that MB may play an important role in the period evolution of sdB + MS systems. In the following sections, we will evolve the synthetic population under various MB laws and compare those results to our observed population.

\subsection{Evolution under magnetic braking}\label{sec: magnetic braking evolution}
We analytically evolved the orbital periods of the PCEBs produced by COSMIC under two MB laws following \citet{el-badry_2022}. Since \citet{el-badry_2022} modeled the evolution of MS + MS binaries in which {\it both} components contribute to the angular momentum loss, we modify their formalism to only include angular momentum losses due to the MS star. Following \citet{el-badry_2022}, we assume that the binary is tidally locked, meaning that the MS star's rotation period is equal to the orbital period, which is consistent with the COSMIC models which assumes tidal synchronization for all Roche lobe overflow binaries.\footnote{In fact, at the longer periods represented in our sample, the sdB is unlikely to be fully synchronized \citep{preece_2018, Ma2024}, but the MS companion is, which is sufficient for MB to operate. When the sdB becomes synchronized at $P_{\rm orb} \approx 6-8$ hours, orbital inspiral will temporarily accelerate as orbital angular momentum is used to spin up the sdB \citep[e.g.][]{Schaffenroth2021}. For a typical system in our sample with $M_{\rm sdB} = 0.5 M_\odot$, $M_{\rm MS} = 0.3 M_\odot$, and $P_{\rm orb} = 8$ hr, we find that the orbital angular momentum exceeds the spin angular momentum of the sdB by a factor of $\sim$1,000. We therefore do not attempt to model the effects of synchronization.} Due to the tidal locking, the angular momentum losses from MB ultimately shrink the orbits. We neglect rotational angular momentum and changes in the component masses and radii during the sdB + MS phase, as well as gravitational wave angular momentum losses. We evolve each binary for a time randomly drawn from $\mathcal{U}(0, 200)$ Myr, which is appropriate for a typical sdB lifetime of $\sim 200$ Myr \citep{Heber_2016} and a constant star formation rate. Lower-mass sdBs have longer helium-burning lifetimes of up to 500 Myr. We show results of the same analysis with an evolution time of $\mathcal{U}(0, 500)$ Myr in Appendix \ref{appendix: longer timelife}. Finally, we draw a random sub sample of the simulated population according to the binaries' eclipse probability (see Appendix \ref{adx: selection function}).  Below, we outline the analysis for the RVJ and the saturated MB prescriptions evolved for $\mathcal{U}(0, 200)$ Myrs.

\subsubsection{RVJ magnetic braking model}
In the RVJ model, the MB torque is given by:
\begin{equation}\label{eq: RVJ torque}
        \dot J_{\rm RVJ} = -a _{\rm RVJ} \left(\frac{M}{M_\odot}\right)\left(\frac{R}{R_\odot}\right)^4\left(\frac{P_{\rm rot}}{1\textrm{d}}\right)^{-3}
\end{equation}
where $a_\textrm{RVJ} \approx 6.8 \times 10^{34}$ ergs. Following \citet{belloni_2023}, we introduce two parameters $k$ and $\eta$ that control the strength of MB above and below the fully-convective boundary:
\begin{equation}\label{eq: RVJ model params}
    \dot{J} = \begin{cases} k \dot{J}_{\rm RVJ}, & M_{MS} > 0.35 M_\odot \\
    \\ \frac{k}{\eta} \dot{J}_{\rm RVJ}, & M_{\rm MS} < 0.35 M_\odot
    \end{cases}
\end{equation}
We describe the analytic period evolution with these modifications in Appendix \ref{adx: rvj model}.

Figure \ref{fig: RVJ model} compares our observed sdB + MS sample to the results of our simulations for populations evolved under the RVJ MB prescription (Equation \ref{eq: RVJ_model}) for a range of $k$ and $\eta$. In each panel, the colored, circular points represent the simulated sdB + MS population evolved under the RVJ MB prescription with the specified $k$ and $\eta$. The zero-age PCEB population was created using COSMIC as described in Section \ref{sec: cosmic}, then evolved and mock-observed using the methods described in Appendix \ref{adx: selection function}. The histograms show the period and companion mass distributions of the simulated and observed populations, normalized to the same scale.

\begin{figure*}[ht!]
\includegraphics[width=\textwidth]{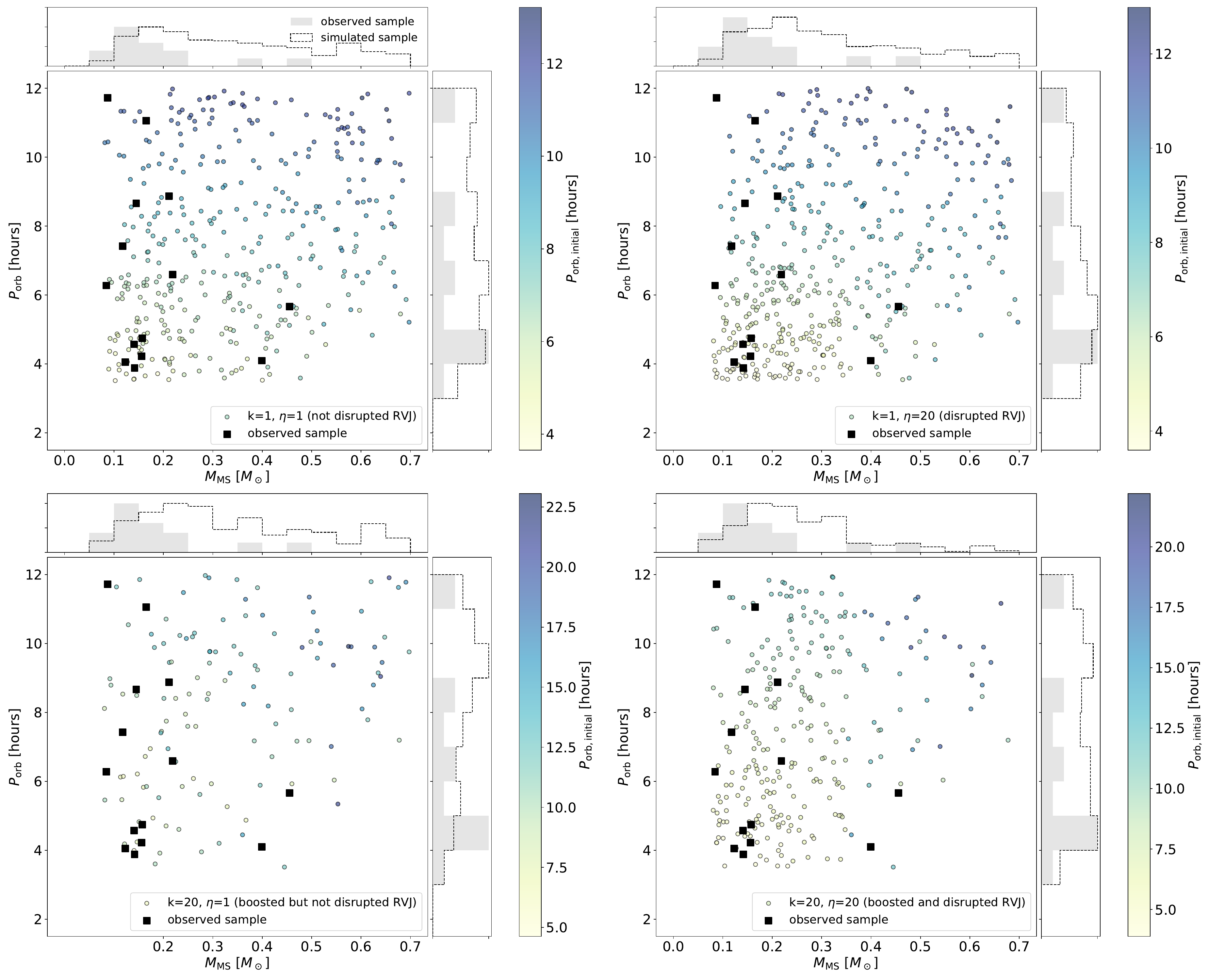}
\caption{Simulations of a population of sdB + MS binaries evolved under the RVJ MB prescription (Equation \ref{eq: RVJ torque}). Black points show our observed sample. Colored points show the result of producing a zero-age PCEB population with COSMIC, evolving their orbital periods for $\mathcal{U}(0,200)$ Myr according to Equation \ref{eq: RVJ_model}, and finally selecting eclipsing systems assuming randomly oriented orbits. All histograms are normalized by their tallest bin. Although this is not representative of the actual bin heights, it allows for a better comparison between the observed and simulated distributions. Each of the panels show the simulated results for various re-scalings $k$ and $\eta$ (see Equation \ref{eq: RVJ model params}). The boosted and disrupted model ($k=20$, $\eta=20$) matches the observed population best. The fiducial RVJ model predicts little period evolution within 100 Myr, and thus too many companions with $M_{\rm MS} > 0.35 M_\odot$.}
\label{fig: RVJ model}
\end{figure*}

The top left panel of Figure \ref{fig: RVJ model} shows the results for an RVJ-like MB prescription with no disruption ($k=1$, $\eta=1$). The simulated population has a fairly uniform companion mass distribution, with a weaker drop-off toward higher companion masses than found in the observed population. We expect the drop in the number of systems above the fully-convective limit to be caused by the disruption in MB; thus, this suggests that our observed population could be reproduced better with a MB prescription with disruption. The top right panel shows the results for a disrupted RVJ prescription ($k=1$, $\eta=20$). Although the simulated companion mass distribution peaks below the fully-convective limit, there are more predicted companions with $M_{\rm MS} > 0.35 M_\odot$ than we find in the observed population. This suggests that if we assume MB is the only process that is removing angular momentum from our systems, the default RVJ prescription (i.e., $k=1$) is not strong enough to produce a sharp  decrease in the number of system above the fully-convective limit. The bottom left panel shows the results for boosted MB without disruption ($k=20$, $\eta=1$). This model again produces a fairly uniform companion mass distribution which suggests that boosting alone does not predict the observed population well. The bottom right panel shows the results for boosted MB with disruption ($k=20$, $\eta=20$). The simulated population best predicts the observed population for this model. 

To assess whether the observed and simulated populations are consistent with one another, we apply a two-sample Kolmogorov–Smirnov (KS) test to their companion mass distributions. We find a p-value of $\sim1\%$ for the boosted and disrupted model with ($k \gtrsim 20$ and $\eta \gtrsim 20$) whereas for the other three models, we find p-values of $0.01-0.05\%$. These values suggests that the observed population is not formally consistent with any of the simulated populations, reflecting the fact that the observed distribution drops off at $M_{\rm MS} \gtrsim 0.2 M_\odot$ instead of $\gtrsim 0.35 M_\odot$. However, the boosted and disrupted RVJ model with $k \gtrsim 20$ and $\eta \gtrsim 20$ best matches the data. The small size of the observed population and imperfect match with simulations prevent us from placing precise constraints on $k$ and $\eta$.

In summary, we find that boosted and disrupted MB model best predict the observed population for the RVJ MB prescription. Here we remain agnostic about the possible physical origins of this stronger MB. We discuss possibilities and the physical plausibility of boosting MB in sdB + MS binaries further in Section \ref{sec: conclusion}.

The period distribution of the simulated population is fairly uniform for all simulations. This can be explained by the large number of surviving systems having a low-mass companion and therefore not experiencing strong MB.

\subsubsection{Saturated magnetic braking model} \label{sec: saturated MB simultion}
\begin{figure*}[ht!]
\includegraphics[width = \textwidth]{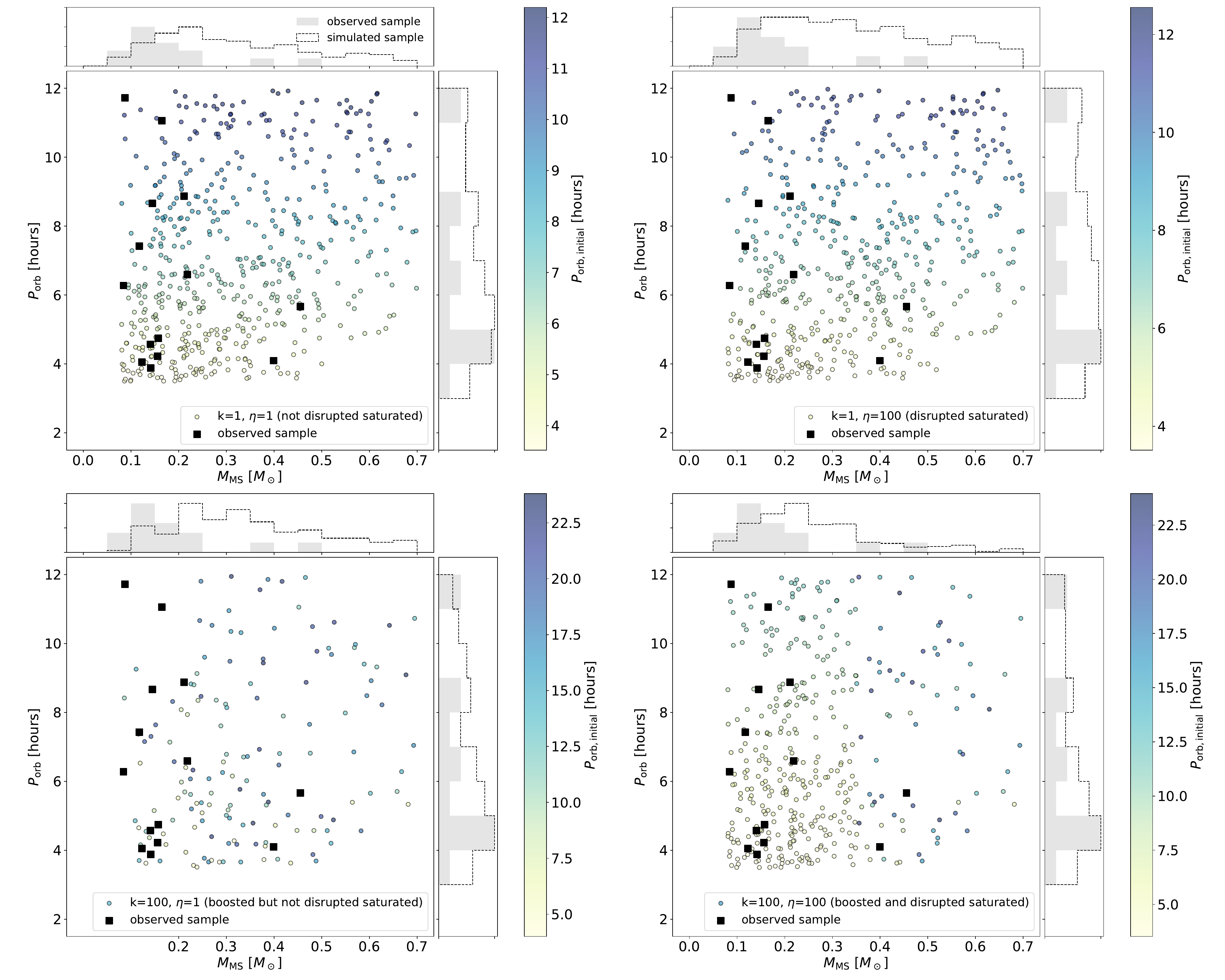}
\caption{Results of our simulations for a population of sdB + MS binaries evolved under the saturated MB prescription (Equation \ref{eq:sat_model}). The boosted and disrupted model ($k=100$, $\eta=100$) matches the observed population best. The fiducial RVJ model predicts little period evolution within 100 Myr, and thus too many companions with $M_{\rm MS} > 0.35 M_\odot$.}
\label{fig: sat model}
\end{figure*}

We next consider a saturated MB law, which predicts a MB torque whose strength scales more weakly with orbital period than the RVJ law \citep[e.g.][]{Kawaler_1988, Chaboyer_1995}. We explore the saturated MB prescription because \citet{el-badry_2022} found that the RVJ MB model predicts a too-steep scaling with $P_{\rm orb}$ at short periods to match the observed MS + MS binary population. Our modeling of the saturated MB law follows \cite{el-badry_2022}. The MB torque is given by:
\begin{equation}\label{eq: sat model}
    \dot{J}_\text{sat} = \begin{cases} 
        -a_\text{sat} \left(\frac{P}{1\,\text{d}}\right)^{-3} \left(\frac{R}{R_\odot}\right)^{1/2} \left(\frac{M}{M_\odot}\right)^{-1/2}, \\
        \qquad \qquad\qquad \qquad\qquad \qquad \qquad\text{if } P \geq P_\text{crit} \\
        -a_\text{sat} \left(\frac{P}{1\,\text{d}}\right)^{-1} \left(\frac{P_\text{crit}}{1\,\text{d}}\right)^{-2} \left(\frac{R}{R_\odot}\right)^{1/2} \left(\frac{M}{M_\odot}\right)^{-1/2},\\
        \qquad \qquad\qquad \qquad\qquad \qquad \qquad\text{if } P < P_\text{crit}
    \end{cases}
\end{equation}
where $a_\textrm{sat} = 1.04 \times 10 ^{35}$ ergs \citep{Sills_2000}. For the period range of interest, all simulated binaries have periods well below the saturation boundary $P_{\rm crit}$; therefore, we use the second equation given in Equation \ref{eq: sat model} for our analysis. Again, we introduce the following parameters $k$ and $\eta$ that control the strength of MB:
\begin{equation}\label{eq: sat model params}
    \dot{J} = \begin{cases} k \dot{J}_{\rm sat}, & M_{MS} > 0.35 M_\odot \\
    \\ \frac{k}{\eta} \dot{J}_{\rm sat}, & M_{\rm MS} < 0.35 M_\odot
    \end{cases}
\end{equation}
We can again solve for an analytical solution (see Appendix \ref{sec: saturated MB}). We repeat the same analysis as in Figure \ref{fig: RVJ model}, now using the saturated MB law. The results of our simulation are compared to the observed sdB + MS population in Figure \ref{fig: sat model}.

Similar to the RVJ prescription, our simulated population best matches our observed population for boosted and disrupted ($k=100$, $\eta=100$) saturated MB. We selected the value of the boost parameter $k$ by comparing the simulated population for various values (i.e., $k = 10, 20, 50, 75, 100$) against the observed sample and choosing the value that best matches the relative drop off we see toward high $M_{\rm MS}$. We chose $\eta$ to be equal to $k$.

We again use a two-sample KS test to compare the observed and simulated populations. We find a p-value of $\sim 1\%$ for the boosted and disrupted model with $k \gtrsim 100$ and $\eta \gtrsim 100$ whereas for the other three models, we find p-values of $0.01-0.05\%$. The small size of our observed sample prevents us from being able to place a tight upper limit on $k$, but we can rule out $k\gtrsim 300$, because in this case the simulations predict too few surviving PCEBs with MS companions above the fully convective limit. Again, here we remain agnostic about the physical origin of this stronger MB (see Section \ref{sec: conclusion} for further discussion). 

The period distributions of the simulated populations for all four variations are fairly uniform, similar to the results from the RVJ MB prescription.

\subsection{Simulating magnetic braking for WD + MS PCEBs}
\begin{figure*}
    \includegraphics[width = \textwidth]{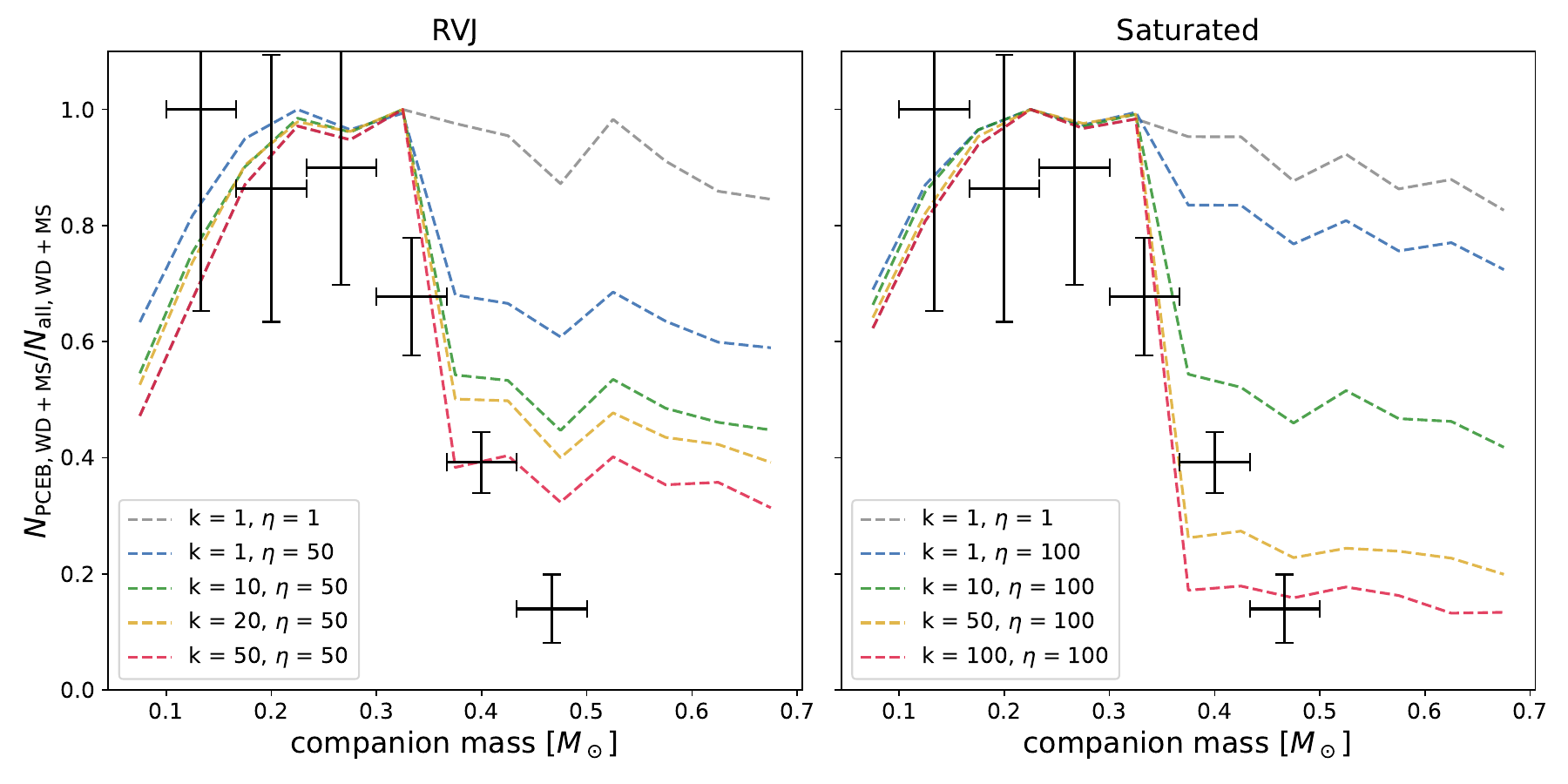}
    \caption{Predicted and observed WD + MS PCEB fractions for a range of MB laws. The observed fractions (black points) are taken from \cite{Schreiber_2010}, while dashed lines show population synthesis models described in the text. Different colors show different values of $k$ and $\eta$. Both the observed and simulated samples are normalized to a maximum value of 1. As for the sdB + MS binaries, we find that -- if the observed dearth of higher-mass companions is a result of MB -- both disruption and boosting of MB above the fully-convective limit are required to match observations.}
    \label{fig: WD+MS simulation}
\end{figure*}
We carried out similar simulations to model the evolution of short-period WD + MS PCEBs. We used the same initial period, eccentricity, and mass distribution models as in Section \ref{sec: cosmic} but instead targeted systems containing He or CO WDs (kstar\_1= 10 or 11), without a cut on WD mass. We considered all binaries that have undergone a common envelope event and applied MB using the same models we used for sdB + MS binaries. Since we expect WD + MS PCEBs to have a typical observable life time of $\sim 1$ Gyr, we choose the evolution time by randomly selecting an age for each system from $\mathcal{U}(0, 2)$ Gyr. We considered as PCEBs all systems with final periods below 4 days, roughly corresponding to the period range below which they would have been recognized as PCEBs with low-resolution spectroscopy \citep{Schreiber_2010}.

Figure \ref{fig: WD+MS simulation} compares our simulated results to the observed WD + MS population from SDSS as reported by \cite{Schreiber_2010}. Following \citet{Schreiber_2010}, we normalize the observed sample of WD + MS population by plotting the ratio of PCEBs to all WD + MS binaries. For the observed sample from SDSS, we take $N_{\rm WD+MS, all}$ to be the number of all systems reported in the \citet{Schreiber_2010} catalogue. For our simulated sample, we take $N_{\rm WD+MS, all}$ to be the number of all WD + MS binaries, including PCEBs and systems that have not undergone a common envelope event. We chose this normalization to keep consistent with \cite{belloni_2023}, who conducted a similar experiment. The left panel of Figure \ref{fig: WD+MS simulation} shows results for the RVJ MB model, and the right panel shows results for the saturated MB model. The different colored dashed lines show predictions for different $k$ and $\eta$. Similarly to the sdB + MS simulations, boosted MB ($k \gtrsim 50$) coupled with strong disruption ($\eta \approx 50-100$) below the fully convective boundary is required to match the observed population. Even for the RVJ prescription, models with disruption and no boosting predict a weaker drop-off above the fully-convective boundary than is observed.

Our results are in agreement with findings from \cite{belloni_2023}, confirming that our modeling of MB and the synthesis of the zero-age PCEBs are consistent with their study.

\section{Summary and Discussion}\label{sec: conclusion}
In order to constrain the companion mass distribution of post-common envelope sdB + MS binaries, we carried out a multi-epoch spectroscopic survey of bright, eclipsing systems. Our main results are as follows:

\begin{enumerate}[label=(\roman*), wide, labelwidth=!, labelindent=0pt]
    \item \textit{Sample selection: }we selected 14 eclipsing sdB + MS systems with $3.8 \leq P_{\rm orb} \leq 12$ hours and apparent magnitude $G < 16$ mag from the list of eclipsing sdB + MS systems reported in \cite{Schaffenroth_2022} (Figure \ref{fig: basic properties}). We measured the orbital periods from their light curves (Figure \ref{fig: lightcurves}). Our sample is $\sim87$\% complete with respect to the \cite{Geier2020} sample in this period and apparent magnitude range and the selection function of our search is straightforward to model. Compared to previous work, our sample contains more long-period binaries (Figure~\ref{fig: population analysis}). This is critical for measuring the companion mass distribution, because shorter-period binaries can only accommodate low-mass companions. 
    
    \item \textit {RV follow-up: } we conducted multi-epoch spectroscopic follow-up to measure the RVs of the sdB stars in our sample over the course of multiple nights (Figure \ref{fig: spectra fits}). The ephemerides measured from the light curves allow us to constrain the sdB RV curves using only a few RV measurements (Figure \ref{fig: radial velocity fits}). Then, assuming a typical mass for the sdB star and an inclination range informed by the fact that the binaries are eclipsing, we constrained the companion masses (Figure \ref{fig: mass function} and \ref{fig: population analysis}).
    \item \textit{Companion mass distribution: } our sample contains companion masses ranging from near the hydrogen burning limit to 0.45 $M_\odot$, including the two most massive MS companions in HW Vir binaries discovered so far. The companion mass distribution peaks near $M_{\rm MS}\sim 0.15 M_\odot$ and exhibits a drop off in a number of systems near $M_{\rm MS}\sim 0.25 -0.30M_\odot$, slightly below the fully-convective limit (Figure \ref{fig: population analysis}). The distribution is similar to what other studies have found for shorter period systems (Figure \ref{fig: population analysis}). We see no correlation between $P_{\rm orb}$ and $M_{\rm MS}$. 
    
    \item \textit{Comparison to binary population models: }we created a synthetic population of zero-age sdB + MS PCEBs using COSMIC. We then evolved the population under various MB prescriptions by using the analytical solution of the period evolution, and mock observed this simulated population. Comparing this simulated sdB + MS population to our observed sample, we find that disrupted of MB, if coupled with boosted MB at higher masses, could explain the observed companion mass distribution (Figure \ref{fig: RVJ model} and \ref{fig: sat model}). This is consistent with the findings by \citet{belloni_2023}, where they conducted a similar experiment with a WD + MS population. There are higher mass WD companions to sdBs in the same period range as our sample, which suggests that \textit{it is unlikely for the observed distribution to be set by common envelope alone} (Figure \ref{fig: population comparison}). 
    \end{enumerate}

Most MB models in the literature, including the RVJ and saturated models we explored, were derived from empirical models of the rotation evolution of single stars. It is possible that MB in close binaries removes more angular momentum than MB in single stars with the same rotation periods. 
 
One possibility is that irradiation of the MS star by its sdB companion boosts MB. Irradiation from the sdB will heat the atmosphere of the MS star, potentially leading to enhanced loss of mass and angular momentum. This may lead to significantly stronger MB in sdB + MS binaries than in single stars or MS + MS binaries, since the MS stars in sdB + MS binaries are more strongly irradiated than those in most other classes of close binaries. However, we also find that some boosting of MB is required to match the companion mass distribution of WD + MS binaries (Figure \ref{fig: WD+MS simulation}), which experience less irradiation.

Various previous works \citep[e.g.][]{RVJ_1983, Spruit_Ritter_1983, Howell_2001, Schreiber_2010, Knigge_2011} have predicted that MB weakens or turns off below the fully-convective boundary. Our data exhibits a drop off in the companion mass distribution near $\sim 0.2 M_\odot$, which could be explained by a disruption in MB for low-mass MS stars. However, comparing our observed sample to the population of WD + MS PCEBs, we find that sdB + MS systems cluster towards lower masses ($M_{\rm MS} \sim 0.1-0.2 M_\odot$) compared to WD + MS systems ($M_{\rm MS} \sim 0.1-0.4 M_\odot$), suggesting that the disruption of MB occurs at different masses for these two binaries (Figure \ref{fig: population comparison}).

If the weakening of MB below the fully convective boundary is gradual rather than abrupt, this might explain why sdB + MS binaries -- which are subject to stronger irradiation-driven mass loss -- contain few MS stars above $\sim0.2 M_\odot$. Given that evidence for disruption is observed at different masses for the two types of PCEBs, this could imply that disruption is driven by binary interactions not accounted for in standard MB models, such as e.g. interaction between the magnetic fields of the two component stars.

Our study is in agreement with findings from \cite{belloni_2023} and provides additional support for boosted and disrupted MB in PCEBs, if we assume that MB is the dominant contributing factor to angular momentum loss and period evolution of PCBEs. However, there is not yet a concrete physical explanation for the boosting and disruption of MB that we infer, and there may be other processes that contribute to the period evolution of these systems. Further investigation is necessary to understand the underling mechanisms that results in this boosting and disruption.

An alternative explanation for the lack of higher-mass companions in post-common envelope sdB binaries could be that more massive companions go through stable mass transfer instead, and thus end up at long orbital periods. This possibility seems unlikely, however, because there are plenty of post-common envelope sdB+WD binaries in short-period orbits containing WDs with masses near $0.6\,M_\odot$. And indeed, most of the known post-stable mass transfer sdB + MS binaries contain MS stars with $M_{\rm MS} \gtrsim 0.8\,M_{\odot}$ \citep[e.g.][]{Vos2018}.


\begin{acknowledgments}

We thank the referee for their constructive comments. We also thank Jim Fuller and Stefan Geier for helpful discussions.  The Kavli Institute for Theoretical Physics (KITP) hosted the program, ``White Dwarfs
as Probes of the Evolution of Planets, Stars, the Milky
Way, and the Expanding Universe,'' during which this
project was initiated.

This research was supported in part by the U.S. National Science Foundation (NSF) grant AST-2307232, and in part by grants PHY-1748958 and AST-2107070.

This work has made use of data from the European Space Agency (ESA) mission
{\it Gaia} (\url{https://www.cosmos.esa.int/gaia}), processed by the {\it Gaia}
Data Processing and Analysis Consortium (DPAC,
\url{https://www.cosmos.esa.int/web/gaia/dpac/consortium}). Funding for the DPAC
has been provided by national institutions, in particular the institutions
participating in the {\it Gaia} Multilateral Agreement.

This work is based in part on observations obtained with the Samuel Oschin 48 inch Telescope at the Palomar Observatory as part of the Zwicky Transient Facility project. ZTF is supported by the NSF
under grant AST-1440341 and a collaboration including Caltech, IPAC, the Weizmann Institute for Science, the Oskar Klein Center at Stockholm University, the University of Maryland, the University of Washington, Deutsches Elektronen-Synchrotron and Humboldt University, Los Alamos National Laboratories, the TANGO Consortium of Taiwan, the University of Wisconsin at Milwaukee, and the Lawrence Berkeley National Laboratory. Operations are conducted by the Caltech Optical Observatories (COO), the Infrared Processing and Analysis Center (IPAC), and
the University of Washington (UW).

Some of the data presented herein were obtained at Keck Observatory, which is a private 501(c)3 non-profit organization operated as a scientific partnership among the California Institute of Technology, the University of California, and the National Aeronautics and Space Administration. The Observatory was made possible by the generous financial support of the W. M. Keck Foundation. 

\end{acknowledgments}

%

\vspace{5mm}


\software{Astropy \citep{astropy:2013, astropy:2018, astropy:2022}, COSMIC \citep{2021ascl.soft08022B}}



\appendix
\section{Radial velocities}\label{adx: RV measurements}

Table of all radial velocities measured by spectroscopic follow-up, their observation time and the instrument used.

\begin{longtable*}[h!]{c c c c} \label{table: follow-up} \\
\hline
object name & observation time (HJD) & radial velocity (km s$^{-1}$) & instrument\\ [0.5ex] 
\hline
J1533+3759 & 2460055.7312 & 60 $\pm$ 10 & DBSP \\
J1533+3759 & 2460055.8064 & -84 $\pm$ 10 & DBSP \\
J1533+3759 & 2460055.8902 & 58 $\pm$ 10 & DBSP \\
J1533+3759 & 2460055.9792 & -69 $\pm$ 10 & DBSP \\
J1533+3759 & 2460344.0607 & -53 $\pm$ 5 & ESI\\
J1533+3759 & 2460344.0649 & -58 $\pm$ 5 & ESI \\
J1533+3759 & 2460344.1242 & 26 $\pm$ 5 & ESI \\
J1533+3759 & 2460344.1553 & 78 $\pm$ 5 & ESI \\
J1920+3722 & 2460056.0040 & -60 $\pm$ 10 & DBSP \\
J1920+3722 & 2460055.9193 & 55 $\pm$ 10 & DBSP \\
J1920+3722 & 2460438.7822 & 85 $\pm$ 10 & DBSP \\
J1920+3722 & 2460463.9611 & 63 $\pm$ 5 & ESI \\
J1920+3722 & 2460464.0117 & -37 $\pm$ 5 & ESI \\
J0557-1409 & 2459905.9428 & 179 $\pm$ 10 & DBSP\\
J0557-1409 & 2459914.0126 & 28 $\pm$ 10 & DBSP\\
J0557-1409 & 2460201.0031 & -89 $\pm$ 10 & DBSP\\
J0557-1409 & 2460202.0067 & -97 $\pm$ 10 & DBSP\\
J0557-1409 & 2460264.9379 & 113 $\pm$ 5 & ESI\\
J0557-1409 & 2460265.0220 & -56 $\pm$ 5 & ESI\\
J0557-1409 & 2460284.8363 & -43 $\pm$ 10 & DBSP\\
J0557-1409 & 2460284.9071 & 107 $\pm$ 10 & DBSP\\
J0619-1417 & 2459906.0087& 0 $\pm$ 10 & DBSP\\
J0619-1417 & 2459914.0063 & 151 $\pm$ 10 & DBSP\\
J0619-1417 & 2460264.9434 & 113 $\pm$ 5 & ESI\\
J0619-1417 & 2460265.0270 & 73 $\pm$ 5 & ESI\\
J0619-1417 & 2460265.0531 & 21 $\pm$ 5 & ESI\\
J0619-1417 & 2460284.8454 & 112 $\pm$ 10 & DBSP\\
J0619-1417 & 2460284.9152 & 58 $\pm$ 10 & DBSP\\
J1852+1445 & 2459899.7008 & 22 $\pm$ 10 & LRIS \\
J1852+1445 & 2459905.5994 & -4 $\pm$ 10 & DBSP \\
J1852+1445 & 2459913.5755 & -40 $\pm$ 10 & DBSP \\
J1852+1445 & 2460055.9488 & 32 $\pm$ 10 & DBSP \\
J1852+1445 & 2460200.6407 & 75 $\pm$ 10 & DBSP \\
J1852+1445 & 2460200.7065 & -32 $\pm$ 10 & DBSP \\
J1852+1445 & 2460225.7755 & 93 $\pm$ 10 & LRIS \\
J1831+1345 & 2459913.5906 & -94 $\pm$ 10 &  DBSP\\
J1831+1345 & 2460055.8568 & -98 $\pm$ 10 &  DBSP\\
J1831+1345 & 2460055.9301 & 46 $\pm$ 10 &  DBSP\\
J1831+1345 & 2460344.1721 & -9 $\pm$ 5 &  ESI\\
J2240+5437 & 2459899.7947 & -155 $\pm$ 10 & LRIS \\
J2240+5437 & 2459899.9316 & 108 $\pm$ 10 & LRIS \\
J2240+5437 & 2460200.6566 & 163 $\pm$ 10 & DBSP \\
J2240+5437 & 2460200.7149 & -34 $\pm$ 10 & DBSP \\
J2240+5437 & 2460284.5866 & 86 $\pm$ 10 & DBSP\\
J0519-1916 & 2459900.0532 & -73 $\pm$ 10 & LRIS \\
J0519-1916 & 2459905.9520  & 78 $\pm$ 10 & DBSP \\
J0519-1916 & 2459915.5922 & 39.48 $\pm$ 5 & FEROS\\
J0519-1916 & 2459916.6030 & 59.40 $\pm$ 5 & FEROS\\
J0519-1916 & 2459917.5771 & -78.84 $\pm$ 5 & FEROS\\
J0519-1916 & 2459919.5742 & -36.88 $\pm$ 5 & FEROS\\
J0519-1916 & 2459920.6471 & -87.63 $\pm$ 5 & FEROS\\
J0519-1916 & 2460264.9327 & -73 $\pm$ 5 & ESI\\
J0519-1916 & 2460265.0728 & 68 $\pm$ 5 & ESI\\
J0519-1916 & 2460284.8283 & 58 $\pm$ 10 & DBSP\\
J0519-1916 & 2460284.9003 & 20 $\pm$ 10 & DBSP\\
J1630+1801 &2460055.8497 & -109 $\pm$ 10 & DBSP\\
J1630+1801 & 2460055.9978 & 2 $\pm$ 10 & DBSP\\
J1630+1801 & 2460344.1483 & -73 $\pm$ 5 & ESI\\
J1802-5532 & 2460374.8937 & -21 $\pm$ 5 & FEROS \\
J1802-5532 & 2460375.8840 & -98 $\pm$ 5 & FEROS \\
J1802-5532 & 2460376.8899 & -104 $\pm$ 5 & FEROS \\
J0808+3202 & 2459905.9611 & 108 $\pm$ 10 & DBSP \\
J0808+3202 & 2459914.0181 & 28 $\pm$ 10 & DBSP \\
J0808+3202 & 2459914.0642 & 114 $\pm$ 10 & DBSP \\
J0808+3202 & 2460201.0083 & 102 $\pm$ 10 & DBSP \\
J0808+3202 & 2460201.0121 & 84 $\pm$ 10 & DBSP \\
J0808+3202 & 2460202.0266 & 48 $\pm$ 10 & DBSP \\
J0808+3202 & 2460264.9680 & 106 $\pm$ 5 & ESI \\
J0808+3202 & 2460265.0583 & 22 $\pm$ 5 & ESI \\
J0808+3202 & 2460343.9595 & -2 $\pm$ 5 & ESI \\
J0808+3202 & 2460344.0580 & 92 $\pm$ 5 & ESI \\
J0808+3202 & 2460344.0538 & 103 $\pm$ 5 & ESI \\
J0241-6855 & 2459912.5673 & 111.00 $\pm$ 5 & FEROS \\
J0241-6855 & 2459912.6816 & 28.26 $\pm$ 5 & FEROS \\
J0241-6855 & 2459914.6449 & -1.35 $\pm$ 5 & FEROS \\
J0241-6855 & 2459914.7741 & 80.22 $\pm$ 5 & FEROS \\
J0612-1740 & 2459914.8046 & -72.09 $\pm$ 5 &  FEROS\\
J0612-1740 & 2459914.6796 & -14.25 $\pm$ 5 &  FEROS\\
J0612-1740 & 2459916.5266 & -9.03 $\pm$ 5 &  FEROS\\
J0612-1740 & 2459917.6901 & -55.11 $\pm$ 5 &  FEROS\\
J0612-1740 & 2459919.5994 & -38.82 $\pm$ 5 &  FEROS\\
J0612-1740 & 2459920.6471 & -68.13 $\pm$ 5 &  FEROS\\
J0612-1740 & 2460284.8519 & -17 $\pm$ 10 &  DBSP\\
J0612-1740 & 2460284.9216 & -82 $\pm$ 10 &  DBSP\\
[1ex] 
\hline \\
\caption{All measured RVs.} \\
\label{table: radial velocities}
\end{longtable*}

\section{Companion masses for high- and low-mass sdBs}\label{appendix: companion mass}
\begin{figure}[ht!]
\begin{center}
\includegraphics[width =0.5\textwidth]{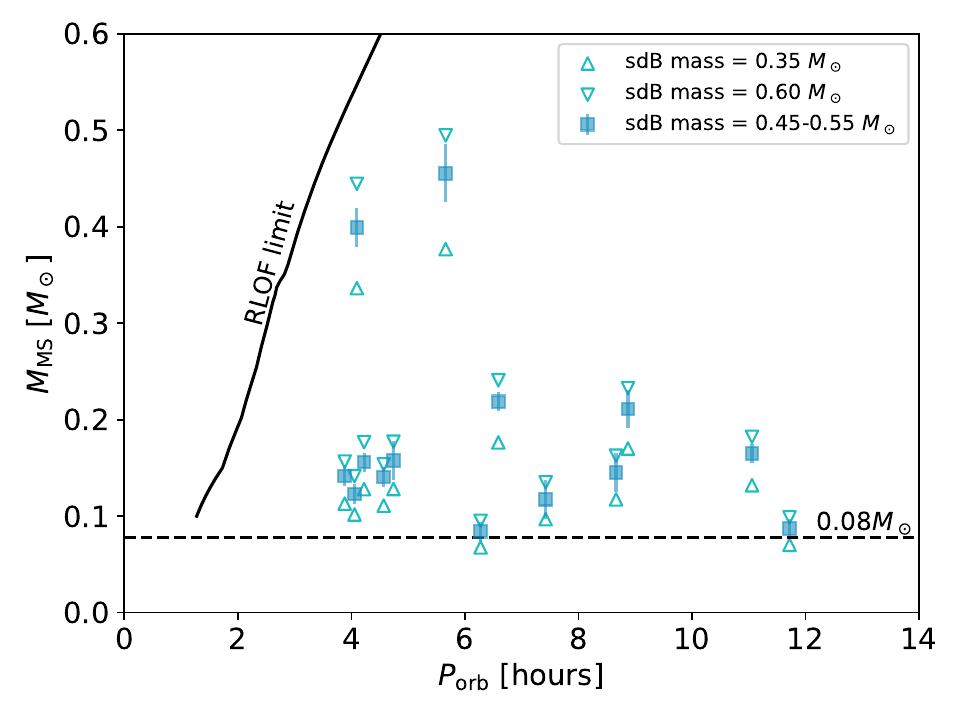}
\end{center}
\caption{Companion mass vs. orbital period. Dashed line marks the hydrogen burning limit; solid line shows where a MS companion would overflow its Roche lobe. Square points with error bars show the fiducial inferred companion masses assuming $M_{\rm sdB} = (0.45-0.55)\,M_\odot$. Triangles show upper and lower limits inferred for $M_{\rm sdB} = 0.60 M_\odot$ and $M_{\rm sdB} = 0.35 M_\odot$, respectively. These more generous limits lead to broader ranges of dynamically-allowed $M_{\rm MS}$ but qualitatively consistent results; e.g., even if all sdBs in our sample had $M_{\rm sdB} = 0.6\,M_{\odot}$, 85\% of the companions would still have $M_{\rm MS}< 0.25\,M_{\odot}$.}
\label{fig: companion mass appendix}
\end{figure}

In our fiducial analysis, we assume sdB masses of $(0.45-0.55)\,M_\odot$. Although we expect the majority of observed sdBs fall in this mass range, theory predicts the existence of both higher-mass ($M_{\rm sdB} \sim 0.60 M_\odot$) and lower-mass ($M_{\rm sdB} \sim 0.35 M_\odot$) sdBs. Here we test the sensitivity of our results to the assumed sdB mass. We repeated the analysis  in Section \ref{sec: companion mass measurement} but assume $M_{\rm sdB} = 0.35 M_\odot$ and $M_{\rm sdB} = 0.60 M_\odot$. The inferred companion mass limits are shown with triangles in Figure \ref{fig: companion mass appendix}. Although these limits span a wider range than the uncertainties reported in Table \ref{table: radial velocity parameters}, the inferred companion mass ranges are still broadly similar to our fiducial constraints and do not move any companions above or below the fully-convective limit. We also stress that we do expect most systems in our sample to have $0.45 < M_{\rm sdB}/M_{\odot} < 0.55$; the limits in Figure~\ref{fig: companion mass appendix} represent extreme cases.

\section{Selection Function}\label{adx: selection function}
Selection function used to mock-observe the simulated population. We impose bounds on the population's orbital period and select based on the eclipse probability of the systems. From geometry of binaries alone, for randomly oriented orbits, the eclipse probability is given by:
\begin{equation}
   p = \frac{R_{\rm sdB}+R_{\rm MS}}{a},
\end{equation}
where $a$ is the semi-major axis of the orbit, and $R_{\textrm{sdB}}$ and $R_{\textrm{ms}}$ are the radii of the sdB and the companion MS star, respectively. 

We use the definition of self-gravitation to define the mass-radius relation for the sdB ($R_{\rm sdB} = \sqrt{GM_{\rm sdB}/g_{\rm sdB}}$) and take a typical value of $\log(g_{\rm sdB} / \textrm{[cm/s}^2]) = 5.5 $. $R_{\rm MS}$ is set by the mass-radius relation from isochrones \citep{MIST}.
\begin{equation}\label{eq: eclipse probability}
    p = \left[\frac{4 \pi^2}{G M_\odot(M_{\text{\rm sdB}}+M_{\text{\rm ms}}){P_{\rm orb}}^2} \right]^{1/3}\left[R_{\rm MS}
+\sqrt{\frac{GM_\odot M_{\rm sdB}}{g_\text{\rm sdB}}} \right]
\end{equation}
With the above eclipse probability, the selection function is given as:

\begin{equation}\label{eq:simulation_cut_criteria}
     \left\{
        \begin{array}{l} 
            3.8 \leq P_{\rm orb} \leq 12 \textrm{ hours} \\
            p = \left[\frac{4 \pi^2}{G M_\odot(M_{\rm sdB}+M_{\rm MS})P^2}\right]^{1/3} \left[R_{\rm MS}
            +\sqrt{\frac{GM_\odot M_{\rm sdB}}{g_{\rm sdB}}} \right] \leq \mathcal{U}(0, 1)
        \end{array}
     \right.
\end{equation}

\section{Magnetic braking analytical models}
We solve for the orbital period evolution of a sdB + MS binary using a similar methods as outlined in \cite{el-badry_2022}. Only the MS star experiences MB, and we assume no mass loss and total tidally locking. 

\subsection{RVJ magnetic braking prescription}\label{adx: rvj model}
The analytical solution for orbital period evolution under the RVJ MB prescription:

\begin{equation}\label{eq: RVJ_model}
     P_{\rm orb}(t)= 
     \begin{cases} 
            \left[ P_{\rm orb, 0}^{10/3} - \frac{k}{\eta}\frac{10 {R_{\rm MS}}^4(q+1)^{1/3}}{q{M_{\rm MS}}^{2/3}} t \right]^{3/10}, & M_{\rm MS} < 0.35 M_\odot \\
            \left[ P_{\rm orb, 0}^{10/3} - k\frac{10{R_{\rm MS}}^4(q+1)^{1/3}}{q{M_{\rm MS}}^{2/3}} t \right]^{3/10}, & M_{\rm MS} > 0.35 M_\odot  
     \end{cases}
\end{equation}
where $P_\textrm{orb, 0}$ is the initial period of the system (i.e., $P_{\rm orb}$ of the zero-age PCEB) in days, $R_{\rm MS}$ is the radius of the companion in solar radii, and $q = M_{\rm sdB}/M_{\rm MS}$. $t$ is a dimension less time defined as $t = T / T_{\rm RVJ}$ where $T$ is the age of the star where we take the age of the star right after the common envelope ejection to be $T=0$, and 
\begin{equation}
    T_{\rm RVJ} = (1\textrm{d})^{1/3}M_\odot^{5/3}\left(\frac{G^2}{2\pi}\right)^{1/3} {a_{\rm RVJ}}^{-1} \approx 5.8  \textrm{Gyrs}.
\end{equation}
Here $k$ and $\eta$ are introduced to allow us to parameterize the ``boost'' and the ``disruption'' of the MB prescriptions as described in the main text of the paper. 

\subsection{saturated magnetic braking prescription}\label{sec: saturated MB}
We carry out a similar calculation for the saturated MB prescription, and solve for the analytical solution of $P_{\rm orb}$:
\begin{equation}\label{eq:sat_model}
     P_{\rm orb}(t)= 
     \begin{cases} 
            \left[ P_\textrm{orb, 0}^{4/3} - \frac{k}{\eta}\frac{4{R_{\rm MS}}^{1/2}(q+1)^{1/3}}{q{M_{\rm MS}}^{13/6}} t \right]^{3/4}, & M_{\rm MS} < 0.35 M_\odot \\
            \left[ P_{\rm orb, 0}^{4/3} - k \frac{4{R_{\rm MS}}^{1/2}(q+1)^{1/3}}{q{M_{\rm MS}}^{13/6}} t \right]^{3/4}, & M_{\rm MS} > 0.35 M_\odot  
     \end{cases}
\end{equation}
$t$ is again a dimensionless time defined as $t = T / T_\textrm{sat}$ where $T_\textrm{sat}$ is defined as:
\begin{equation}
    T_{\rm sat} = \frac{(1 \textrm{d})^{4/3} M_\odot^{5/3}G^{2/3}}{(2\pi)^{4/3}K_W\omega_{\rm crit, 1}^2}
\end{equation}
where $K_W = 2.7 \times 10^{47}$ g cm$^{2}$ \citep{Sills_2000}. 

\section{variations on the magnetic braking simulations}

\subsection{PCEBs hosting low-mass sdBs}\label{appendix: COSMIC larger mass range}

\begin{figure}[h!]
    \centering
    \includegraphics[width=0.5\linewidth]{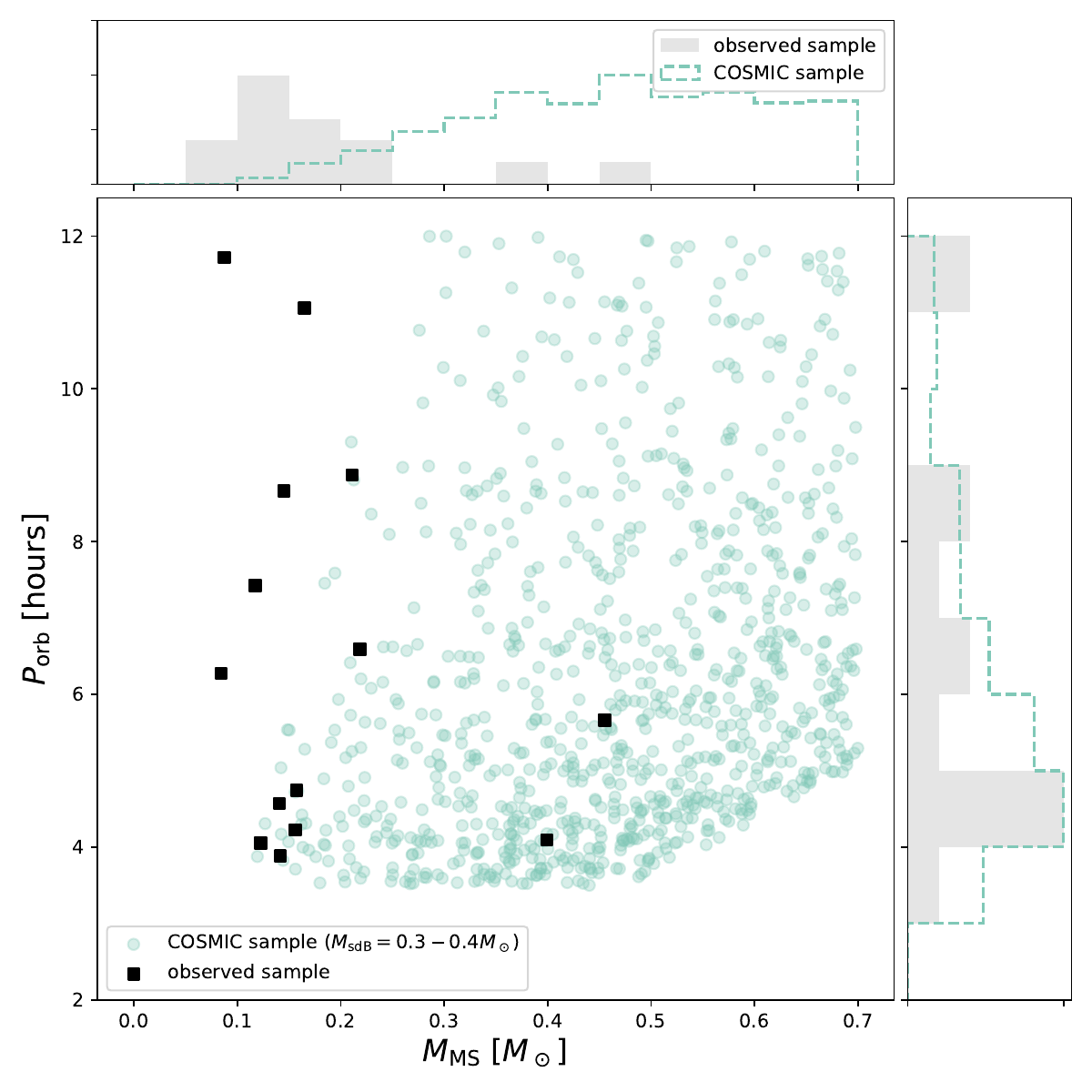}
    \caption{Synthetic zero-age sdB + MS PCEB population predicted by COSMIC for binaries with WD mass $M_{\rm WD} = 0.3-0.4 M_\odot$ (compare to Figure~\ref{fig: cosmic_population}). Black squares show our observed sample. The predicted PCEB population is heavily biased towards higher-mass companions, because would-be PCEBs with lower-mass companions are predicted to merge.}
    \label{fig: COSMIC larger mass range}
\end{figure}

Our analysis in Section~\ref{sec: comparing observations to simulated populations} selects binaries hosting $0.4-0.5\,M_{\odot}$ He WDs as representative of sdBs formed through the canonical channel. This selection is likely inappropriate for lower-mass sdBs formed from $\sim 2\,M_{\odot}$ progenitors that ignited helium burning when their cores were non-degenerate \citep[e.g.][]{Han_2002}. We investigate predictions for lower-mass sdBs in Figure~\ref{fig: COSMIC larger mass range}, where we compare the observed population to COSMIC simulations with initial $M_1 > 2\,M_{\odot}$ that form He WDs with mass $0.30 - 0.40 M_\odot$. Compared to our fiducial simulations, the PCEB population predicted in this case is heavily biased towards high-mass companions, with few binaries predicted to have $M_{\rm MS} < 0.15\,M_{\odot}$ and less overlap with the observed population. COSMIC predicts low-mass companions that enter a common envelope with a star low on the giant branch -- as is required to form a low-mass WD or sdB -- will not survive, because they lack the orbital energy required to eject the giant's envelope. For this reason, we restrict our analysis in the main text to PCEBs with $0.4-0.5 M_\odot$ He WDs.

\subsection{Sensitivity to the assumed sdB lifetime} \label{appendix: longer timelife}

\begin{figure}[h!]
\includegraphics[width = \textwidth]{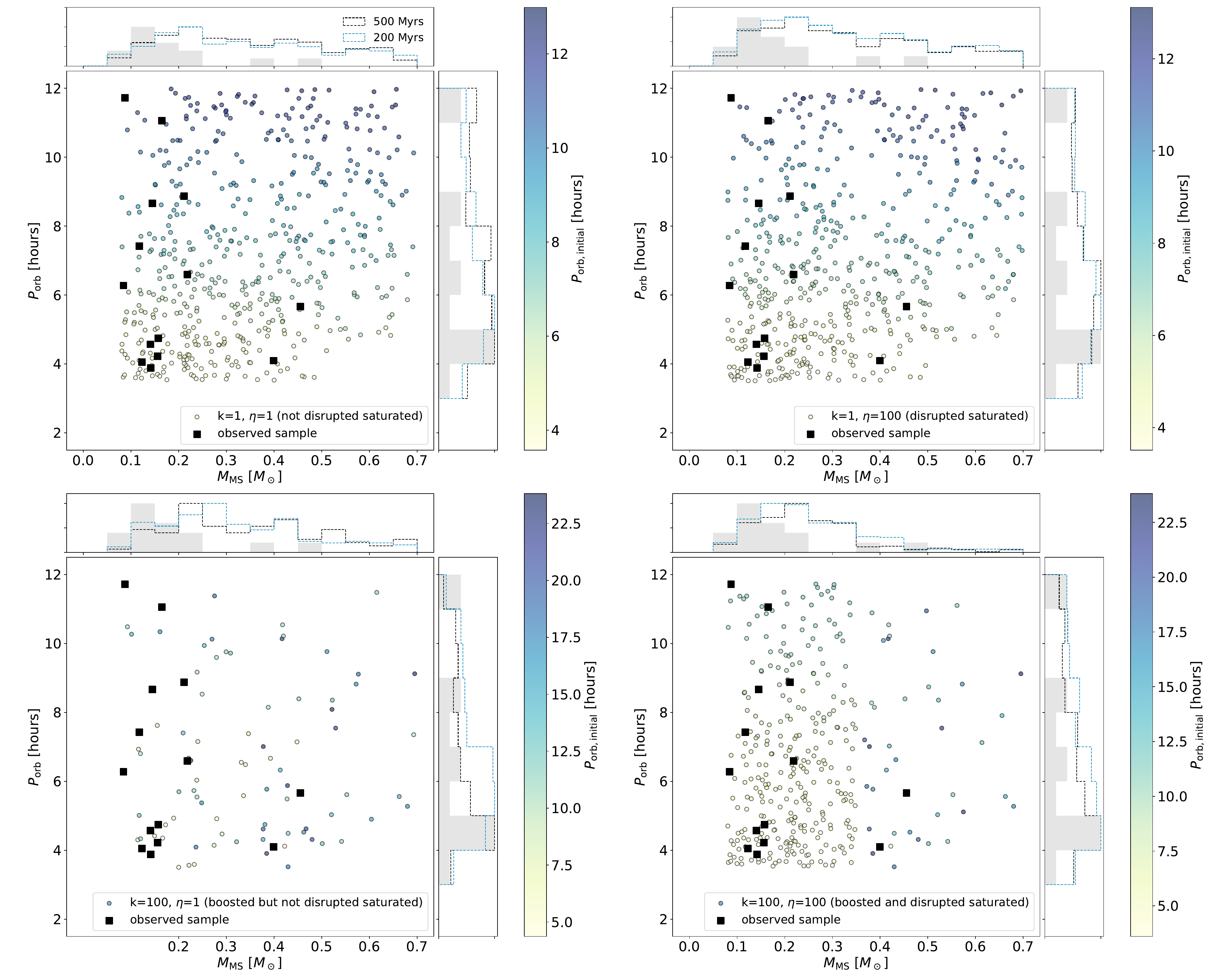}
\caption{Simulations of a population of sdB + MS binaries evolved under the saturated MB prescription (Equation \ref{eq: RVJ torque}). Black points show our observed sample. Colored points show a predicted PCEB population from COSMIC evolved for $\mathcal{U}(0,500)$ Myr according to Equation \ref{eq: sat model} and selected to be  eclipsing. The dotted black line in the histogram represents the population which has been evolved for $\mathcal{U}(0,500)$ Myr and the blue line represents the population which has been evolved for $\mathcal{U}(0,200)$ Myr, as assumed in our fiducial model. All histograms are normalized by their tallest bin. There is little difference in the companion mass distributions predicted between the two age distributions.}
\label{fig: saturated simulation appendix}
\end{figure}

In Section \ref{sec: saturated MB simultion}, we assume that  sdBs have a lifetime of $200$ Myr. Although this is reasonable for typical sdBs, the He burning lifetime of lower-mass sdBs ($M_{\rm sdB} \sim 0.35 M_\odot$) can be longer, up to 500 Myr. To test the sensitivity of our results to the assumed sdB lifetime, we repeat the same analysis but with an evolution time drawn from $\mathcal{U}(0,500)$ Myr. The results are shown in Figure \ref{fig: saturated simulation appendix}. The population evolved for $\mathcal{U}(0,200)$ Myr and $\mathcal{U}(0,500)$ Myr have similar companion mass distributions, suggesting the simulations are insensitive to the assumed sdB lifetime.

\end{document}